\documentclass[10pt,conference]{IEEEtran}

\usepackage{amsmath}
\usepackage{amssymb}
\usepackage{booktabs}
\usepackage{enumitem}
\usepackage{hyperref}
\usepackage{xcolor}
\usepackage{tikz}
\usepackage{graphicx}
\usepackage{eso-pic}
\usetikzlibrary{arrows.meta,positioning,shapes.geometric,fit,calc}

\hypersetup{
  colorlinks=true,
  linkcolor=blue,
  citecolor=blue,
  urlcolor=blue
}

\title{OpenAgenet / OAN White Paper: Open Infrastructure for Trusted Agent Interconnection}

\author{
\IEEEauthorblockN{Jinliang Xu}
\IEEEauthorblockA{\textit{China Academy of Information and Communications Technology}, Beijing, China \\
xujinliang@caict.ac.cn; jlxufly@gmail.com}
}

\begin{document}

\AddToShipoutPictureBG*{%
  \begin{tikzpicture}[remember picture,overlay]
    \node[opacity=0.22] at (current page.center) {%
      \includegraphics[width=.333\paperwidth,keepaspectratio]{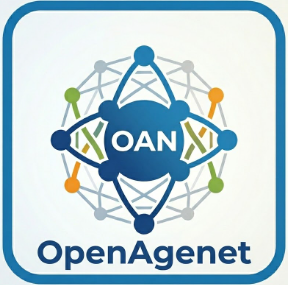}%
    };
  \end{tikzpicture}%
}

\maketitle

\begin{abstract}
OpenAgenet (OAN) is an open infrastructure project for trusted Agent
interconnection across independently operated networks. It addresses the
verification of resource identity, governance state, discovery authorization,
and freshness before interaction. OAN combines decentralized identifiers
(DIDs), verifiable credentials (VCs), and on-chain infrastructure governance
with a routing-aware \texttt{did:oan} registration and discovery model. Registrars verify
onboarding evidence; the Root service accepts resource identity versions and
publishes verifiable packages; authorized Discovery nodes return signed
results within their governed scope. Root also governs the semantic tag tree
shared by registration and discovery. Public DID Documents and credentials
remain separate from local OAN Identity signing material and from controller
authorization. Skill, Runtime, Sidecar, and Gateway integration paths allow
existing Agents to adopt this identity model progressively, including through
the Model Context Protocol (MCP) and Agent-to-Agent (A2A) protocol, while
retaining their native task and workflow semantics. Prototype evaluation and a
public pilot support examination of the trust lifecycle and its operational
bottlenecks. This white paper presents the architecture, governance model,
ecosystem adoption paths, and standardization opportunities for OAN as a shared
identity and trust infrastructure for heterogeneous Agent resources.
\end{abstract}

\noindent\textbf{Public portal:}
\href{https://www.openagenet.xyz/}{https://www.openagenet.xyz/}

\noindent\textbf{Open-source project:}
\href{https://github.com/OpenAgenet}{https://github.com/OpenAgenet}


\section{Executive Summary}

Large language model (LLM)-based Agents are changing from local assistants into
network participants. They call tools, consume data, expose capabilities,
delegate work, negotiate task context, and cooperate with other Agents. As this
shift continues, message exchange depends on a prior infrastructure task:
verifying a discovered Agent's identity, the authority that accepted it, its
current status, and the Discovery service's authority to expose it. A relying
Agent needs machine-verifiable evidence for these decisions before interaction.

OAN addresses this pre-connection trust problem through routing-aware
\texttt{did:oan} registration, governance-backed Root acceptance, verifiable
package distribution, authorization-aware Discovery, controller authorization,
and signed invocation. These mechanisms establish identity, lifecycle, and
provenance evidence while leaving task semantics, tool protocols, model
orchestration, dialogue strategy, and application workflow to higher-layer
protocols and frameworks. Existing Agent systems can adopt this infrastructure
incrementally, from governed publication to evidence-aware MCP or A2A
interaction through a local runtime or deployment boundary.

\begin{figure}[!ht]
\centering
\includegraphics[width=\columnwidth]{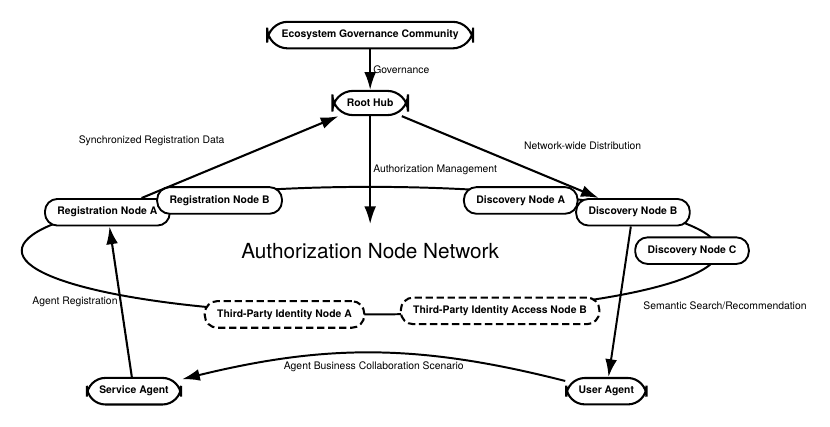}
\caption{OpenAgenet / OAN framework overview. OAN separates governance, Root
acceptance, Registrar onboarding, Discovery exposure, VC issuance, and verified
invocation into distinct roles within an authorized node network.}
\label{fig:wp-oan-framework}
\end{figure}

Figure~\ref{fig:wp-oan-framework} provides a compact orientation for the
governance, Root, Registrar, Discovery, and interaction roles used throughout
the paper.

OAN separates three identity artifacts that are often conflated. The DID
Document is the public document associated with a decentralized identifier
(DID).
OAN Identity is local private identity material that binds a DID, a DID
Document, public key material, and
private key material for signing. A verifiable credential (VC) is a signed
issuer statement about an authorization or registration fact. This separation allows a resource to be
self-controlled, controlled by a dedicated controller, or controlled by an
external controller with explicit authorization proof, without turning the
controller string itself into a trust shortcut.

The same separation of public evidence and local keys allows OAN identities
to support application authentication as well as Agent interaction, while
keeping private signing material under the holder's control.

Infrastructure legitimacy is determined through governance state and Root
verification. Infrastructure lifecycle decisions for Registrars,
Discovery nodes, and future VC issuers are published by an on-chain governance
layer that can be operated under committee and threshold rules. Root consumes
that governance state, enforces it in service admission, and converts approved
state into protocol-level authorization VCs. Root is therefore closer to a
policy enforcement and credential issuance point than to the ultimate social
governance authority.

OAN is a \textbf{trusted resource registry and discovery infrastructure}.
Unlike registries centered on one object class, it applies the same verifiable
identity, lifecycle, package, and discovery path to Agent Services, Skills, MCP
Servers, and Tool/application programming interface (API) endpoints. This
resource-oriented model can also
accommodate later object classes, such as workflows, datasets, or model
endpoints, without assigning OAN responsibility for their native execution.

This resource model has implications for standardization. If the ecosystem standardizes
only Agent records, it may later need separate registries for Skills, tools,
model endpoints, and workflow assets. If it standardizes only tool registries,
it cannot express lifecycle governance for non-tool resources. OAN therefore
places the ResourcePackage at the center: a resource can keep its native
artifact or protocol, while OAN standardizes the identity, provenance,
registration evidence, lifecycle status, package proof, and discovery evidence
around it. This is similar in spirit to artifact distribution systems such as
the Open Container Initiative (OCI), where different artifacts can be distributed through a common manifest and
registry discipline~\cite{oci_distribution_spec,oci_image_spec}; OAN applies a
trust-governed version of that idea to Internet of Agents resources.

The project is motivated by a practical observation. A future Agent ecosystem
will not be operated by one platform. Agents will be developed by enterprises,
research institutes, individual developers, public-service systems, industrial
platforms, and vertical-domain operators. They will run behind different
gateways, use different runtimes, and support different interaction protocols.
If every operator defines its own identity admission, directory, credential,
and trust-verification mechanism, the ecosystem fragments quickly. OAN provides
a common trust substrate that allows heterogeneous Agent Services, Skills, MCP
Servers, and Tool/API resources to become discoverable and verifiable under
shared governance rules.

OAN follows five design principles:

\begin{itemize}[leftmargin=*]
  \item \textbf{Identity before interaction}: a discoverable resource should
  have a governed digital identity before it becomes visible to other parties.
  \item \textbf{Lifecycle before directory}: Discovery should expose only
  Root-accepted, current, and policy-admissible identity versions.
  \item \textbf{Authorization before exposure}: a Discovery service should
  only expose resource identities within its authorized domains.
  \item \textbf{Verification before invocation}: a relying party should verify
  discovery provenance and request signatures before business interaction.
  \item \textbf{Compatibility before replacement}: OAN should complement MCP,
  A2A, Agent Network Protocol (ANP), DID/VC methods, and domain protocols
  instead of replacing them.
\end{itemize}

The result is a governance-oriented Agent interconnection layer. It gives
different organizations a way to cooperate without forcing them into one Agent
runtime, one model vendor, one application framework, or one task protocol.
The architectural contribution is the coupling of federated infrastructure
governance, resource identity, verifiable distribution, authorization-aware
Discovery, and pre-connection verification.
This structure supports heterogeneous resource publication, DID-based semantic
discovery, blockchain-backed federated governance, and integration through
software development kits (SDKs) and reusable identity components. Resource
providers, consumers, and infrastructure operators participate through shared
verification rules while retaining their own application environments.

OAN is therefore more than a search layer. Discovery identifies candidates;
Root records and VCs provide evidence for admission decisions; and the
ResourcePackage binds accepted identity state to a verifiable distribution
object. Root also maintains the governed tag tree used by Registrars and
Discovery nodes, while application protocols remain responsible for task
execution.

This role separation supports integration with specialized communities.
Schema communities can improve how resources describe
themselves. Directory communities can improve routing, indexing, and scale.
Naming communities can improve human-friendly aliases. Runtime communities can
improve execution. OAN's contribution is the trust envelope under which those
outputs can be registered, distributed, discovered, and verified under shared
trust rules.

\subsection{Contributions and Reading Guide}

The white paper makes three connected contributions. It defines a
resource-oriented identity model in which Agent Services, Skills, MCP Servers,
and Tool/API endpoints share a verifiable registration and discovery
lifecycle. It separates the evidence and responsibilities of governance,
Root, Registrar, CDN, Discovery, resource controllers, and relying Agents, so
that no directory response or controller string is treated as sufficient trust
on its own. It also describes a staged adoption path: an existing system can
publish and discover OAN identities first, then add evidence-aware interaction
and identity-bound MCP or A2A exchanges through the SDK, Runtime, Sidecar, or
Gateway interfaces.

The paper first develops the motivation, scope, and design principles, then
defines the system roles, trust architecture, and identity adoption paths.
Subsequent sections examine authorization, protocol relationships, deployment,
governance, security, and evaluation before addressing standardization and
ecosystem participation. The technical yellow paper, protocol
specifications, and conformance artifacts should be used for wire-level
formats, state machines, validation algorithms, and implementation details.

\section{Background: From Local Agents to Open Agent Networks}

Early Agent systems were often embedded inside one application. The application
owner controlled the model, prompt, tool list, user interface, permissions, and
execution environment. In that setting, identity and discovery were simple
because the platform boundary was clear. The application already knew which
components it trusted.

Open Agent interconnection changes this assumption. A User Agent may need to
find a professional Service Agent operated by another organization. A Service
Agent may rely on tools exposed through MCP servers. A workflow platform may
hand a task to an A2A-compatible Agent. A public-domain directory may publish
Agents that serve public services, industry data, or enterprise operations.
Recent multi-agent systems and surveys show the same shift from isolated Agent
prototypes toward networked, tool-using, and collaborative Agent
workflows~\cite{wang2024llmagentsurvey,wu2023autogen,hong2024metagpt,park2023generativeagents}.
This open setting creates four infrastructure gaps.

\subsection{Identity Gap}

An Agent needs an identity that is more than a display name, endpoint URL, or
marketing description. It needs verifiable subject information, verification
methods, service endpoints, capability metadata, and lifecycle state. The
resource identity should be signed, versioned, and traceable to a governance
decision.
Without such a representation, relying parties cannot distinguish an accepted
Agent from an imitation, stale copy, or unreviewed endpoint.

\subsection{Governance Gap}

Open networks need a way to decide which infrastructure participants can
register Agents, which Discovery services can expose which domains, and which
resource identity versions are accepted. Governance is not only a social process.
It must be reflected in machine-verifiable records. Otherwise, operators must
manually trust directory contents and cannot automate validation.

\subsection{Discovery Gap}

Discovery is not merely search. In an open Agent network, Discovery is a
controlled exposure function. It should return candidates that match the query
and also satisfy identity provenance, freshness, and authorization constraints.
A Discovery node that is allowed to expose manufacturing Agents should not
automatically expose medical, financial, or government-service Agents simply
because it can fetch their metadata.

\subsection{Invocation Gap}

Even after discovery, a Service Agent should not accept every incoming request
that names a known Agent. The request should be bound to caller identity, target
identity, body hash, timestamp, nonce, and proof. This guard is especially
important when Agents trigger actions, retrieve data, or initiate downstream
tool calls.

OAN is designed to close these gaps while remaining small enough to be
implemented by independent operators.

\section{Why OAN, Why Now}

The Agent ecosystem is entering a phase where interaction protocols are moving
faster than trust infrastructure. MCP makes tools and data sources easier for
AI applications to access~\cite{mcp_spec}. A2A-style protocols make
Agent-to-Agent task exchange more explicit~\cite{a2a_spec}. Agent DID methods
and ANP-like efforts explore identity and networking
conventions~\cite{did_wba_spec}. These developments are necessary, but they
leave a common precondition unresolved: before an Agent enters a tool session,
task session, or domain workflow, the relying party must know whether the
Agent identity is governed, current, discoverable under policy, and verifiable.

Traditional approaches are not sufficient by themselves. API keys identify
access relationships but do not provide open discovery or lifecycle provenance.
Private directories work inside one platform but do not scale across
organizations. Ordinary service registries expose endpoints but usually do not
encode Agent subject control, governance acceptance, or Discovery authorization.
DID documents and Verifiable Credentials provide important identifier and
credential substrates~\cite{w3c_did_core,vc_data_model_2}, but they do not by
themselves define who may register an Agent into an ecosystem, which Discovery
node may expose it, or how a relying Agent should validate pre-connection
evidence.

OAN is timely because open Agent cooperation needs a layer that is neither a
closed platform account system nor another task protocol. It is a trust
infrastructure layer that can sit beside existing protocols and make them safer
to use across organizational boundaries.

The problem is no longer limited to Agent service endpoints. Agent ecosystems
are also beginning to publish Skills, MCP Servers, Tool/API endpoints,
workflow components, prompts, datasets, and model-facing services. These
resources are usually scattered across code repositories, portals,
marketplaces, enterprise systems, and protocol-specific directories. A normal
search engine or endpoint list can help a user find where something is, but it
cannot by itself answer whether the resource is current, whether the publisher
controls the claimed identity, whether a governance process accepted it, or
whether a Discovery node is authorized to expose it. OAN turns this repeated
application-level problem into shared trust infrastructure.

This shift also changes the unit of trust. In many early Agent systems, the
main question was whether one Agent endpoint could be called. In open
ecosystems, the harder question is whether many heterogeneous resources,
operated by different organizations and discovered through different nodes, can
share a common pre-connection trust path. OAN therefore treats trusted
interconnection as a resource-network problem: resources need identities,
governance evidence, lifecycle state, semantic visibility, and verification
material before task-level protocols begin.

\section{Vision and Scope}

The long-term vision of OAN is an open Internet of Agents in which Agents and other
AI resource products can be registered, discovered, verified, and invoked
across organizational boundaries under common trust rules. Each trust domain
combines a governance layer, a Root service, and authorized infrastructure nodes.
Committee-governed infrastructure lifecycle decisions support multi-operator
participation within a domain, while cross-domain federation profiles can build
on these foundations to connect independently governed domains. OAN provides a
governed trust-domain model, an end-to-end identity lifecycle, multiple Registrar
and Discovery roles, and integration interfaces for existing Agents. Together,
these elements support an open ecosystem that can expand without imposing one
global governance authority or a single Agent protocol.

\subsection{What OAN Is}

OAN is:

\begin{itemize}[leftmargin=*]
  \item a lifecycle model for governed AI resource identity;
  \item a resource-oriented DID method for Agent Services, Skills, MCP
  Servers, and Tool/API endpoints;
  \item a federated infrastructure governance model for Root, Registrar,
  Discovery, and future VC issuer roles;
  \item a package-distribution model for Root-verified identity artifacts;
  \item an authorization-aware Discovery model based on authorized domains;
  \item a pre-connection verification model for Agent-to-Agent invocation;
  \item a public pilot surface for observing, registering, discovering, and
  evaluating trusted resources;
  \item a reference implementation profile and developer-facing integration
  surfaces for ecosystem participation.
\end{itemize}

\subsection{What OAN Is Not}

OAN is not:

\begin{itemize}[leftmargin=*]
  \item a replacement for MCP, A2A, ANP, or other interaction protocols;
  \item a model runtime, LLM inference framework, or Agent-planning engine;
  \item a ranking algorithm or marketplace business model;
  \item a universal semantic ontology for all Agent capabilities;
  \item a centralized platform that must observe all Agent business traffic.
\end{itemize}

OAN is positioned below application
protocols and above raw cryptographic primitives. It gives other protocols a
shared trust context, but it does not prescribe their internal task semantics.
This white paper therefore does not freeze wire formats, API schemas,
canonical serialization, or SDK function signatures. Those implementation
details belong in the technical yellow paper, protocol specifications, and
conformance artifacts. The purpose here is to define the value, roles, trust
path, governance vocabulary, and adoption model that make those lower-level
specifications meaningful.

\section{Design Principles}

\subsection{Governance and Protocol Should Meet at Verifiable Facts}

Real ecosystems require governance. Operators must authorize infrastructure
nodes, review participants, define authorized domains, handle revocation, and
audit misbehavior. At the same time, runtime systems require machine-verifiable
facts. OAN connects these two sides through signed records, Root-verified
packages, bulletin facts, credentials, document hashes, and signed responses.

\subsection{Discovery Should Be Policy-Aware}

Traditional service discovery often focuses on name, address, or metadata
matching~\cite{cheshire2013dns,rfc9460}. OAN treats discovery as an
authorization-sensitive function. A candidate is not admissible merely because
it exists and matches a keyword.
It must be accepted by Root, current, valid under authorized-domain policy, and
returned by an authorized Discovery node.

\subsection{Compatibility Should Be Designed In}

The Agent ecosystem already contains tool protocols, task protocols, DID
methods, API gateways, service meshes, and enterprise identity systems. OAN
should not require these systems to be discarded. It should provide adapters,
metadata mapping, and pre-connection checks that allow existing protocols to
benefit from governed identity and discovery.

\subsection{Trust Should Be Distributed by Role}

No single operational component should silently inherit all trust. The
governance layer records infrastructure lifecycle decisions; Root enforces
those decisions and issues protocol credentials; Registrar helps with
onboarding but does not decide final acceptance; the content delivery network
(CDN) distributes packages but
does not become a trust authority; Discovery queries and indexes data but must
verify Root proof and respect authorization. Service Agents execute business
logic but should verify callers and provenance. This role separation reduces
implicit trust and creates clearer audit boundaries.

\subsection{The Architecture Should Be Observable and Testable}

The reference system turns OAN's trust architecture into an executable,
observable network across lifecycle, publication, synchronization, and query
paths. Its evaluation validates these mechanisms and provides concrete feedback
for scaling them across operators and deployment environments.

\section{System Roles}

OAN separates infrastructure roles from discoverable resource products. Each
infrastructure role can be operated by one organization in a small pilot or by
different organizations in a larger trust domain. Discoverable products can be
published by Agent developers, enterprises, research groups, tool providers,
or community maintainers.

\subsection{Root Node}

The Root Node is the policy enforcement, verification, and credential issuance
center of one OAN trust domain. It observes infrastructure lifecycle state from
the on-chain governance layer, issues Root-signed infrastructure authorization
VCs, verifies complete resource identity submissions, anchors accepted
versions, manages authorization-domain governance, publishes Root-signed records,
and creates verified packages for distribution. Root is not expected to handle
ordinary business traffic. Its purpose is to establish which resource
identities are admissible and to enforce governance-derived infrastructure
authorization in protocol interactions.

Root has three hub functions in the trust domain. It is the
\emph{trust authorization hub} because it accepts resource versions, issues
infrastructure credentials, and signs protocol evidence. It is the
\emph{data-distribution hub} because it
turns accepted identity state into Root-verified packages and publication
cursors consumed by distribution and Discovery services. It is also the
\emph{semantic-governance hub} because it governs the ecosystem semantic tag
tree used by Registrars during label selection and recommendation and by
Discovery nodes during semantic discovery. Root does not perform every search
query, but it governs the shared semantic vocabulary that makes cross-node
Discovery comparable.

Root maintains the following categories of state:

\begin{itemize}[leftmargin=*]
  \item active and inactive Registrar authorization state;
  \item active and inactive Discovery authorization state;
  \item future third-party VC issuer authorization state;
  \item Root-issued infrastructure authorization VCs;
  \item accepted resource identity versions;
  \item rejected submissions and verification reasons;
  \item authorization-domain governance metadata;
  \item semantic tag tree versions and governance metadata;
  \item publication cursors, package metadata, and bulletin facts.
\end{itemize}

Root is intentionally conservative. It should reject malformed \texttt{did:oan}
DID documents, invalid registration credentials, unauthorized Registrar
submissions, stale or tampered representations, and authorization-domain
inconsistencies.

In this governance model, Root does not treat the on-chain governance
layer as a credential. The governance layer records lifecycle state, while the
Root-issued VC is the service credential used in protocol interactions.
Effective infrastructure authorization requires both: active governance state
and a valid Root-issued VC. This keeps Root as the verification and credential
issuer without making Root the only status-query hub for every infrastructure
node.

\subsection{Registrar Node}

The Registrar Node is the onboarding gateway. It helps resource providers
prepare identity drafts, select capability tags, prove control of DID
verification methods, receive registration credentials, and submit complete
identity representations to Root. Registrar improves usability while preserving
Root as the final resource-acceptance and package-verification authority inside
the trust domain.

Registrar binds the onboarding workflow to
protocol evidence. In the reference design, a Registrar draft can move through
creation, product-type declaration, semantic description, capability tagging,
DID-control challenge, credential issuance, and Root submission. The credential
issued by Registrar allows Root to verify that the submitted resource identity
passed the expected onboarding path.

\subsection{Discovery Node}

The Discovery Node is a controlled query service. It synchronizes Root-verified
packages, verifies Root proof, checks bulletin facts, applies authorized-domain
authorization, builds local semantic indexes, and returns signed discovery
responses. It does not decide that a resource identity is accepted; it relies
on Root proof. It also does not expose every package it can fetch. Its own
authorization state limits which domains it may index and expose.

Discovery may add ranking, search optimization, local indexes, reputation
signals, or domain-specific filters, but these should operate after basic OAN
admissibility checks.

For semantic discovery, OAN can draw on the GRAIL line of work, which studies
deep-granularity hybrid resonance and small language model (SLM)-enhanced indexing for real-time Agent
discovery~\cite{xu2026grail}. In OAN, such semantic matching is treated as a
Discovery-node capability: it improves recall, ranking, and intent matching
over verified resource metadata, but it does not replace Root acceptance,
package proof, or Discovery authorization checks.

\subsection{CDN Service}

The CDN Service distributes Root-verified packages, manifests, metadata, and
documents. It improves availability and reduces load on Root. It is not a
trust authority. A Discovery node or relying party should verify hashes,
signatures, package metadata, and Root proof rather than trusting the CDN path
alone.

\subsection{Discoverable Resource Products}

OAN supports several resource products under one identity and discovery model:

\begin{itemize}[leftmargin=*]
  \item \textbf{Agent Service}: a network-accessible Agent that exposes callable
  capabilities and may support A2A, MCP-adjacent, ANP-like, or domain APIs.
  \item \textbf{Skill}: a portable capability description or skill artifact
  whose DID Document provides semantic discovery fields and may reference the
  external artifact location.
  \item \textbf{MCP Server}: a tool and resource server whose endpoint and
  protocol metadata can be bound to a governed \texttt{did:oan} identity before
  MCP initialization.
  \item \textbf{Tool/API}: a conventional API or tool endpoint that benefits
  from OAN registration, discovery, and credential-bound trust verification.
\end{itemize}

Each product can be represented by a
\texttt{did:oan} DID Document with product type, semantic labels, descriptions,
use cases, endpoints or artifact references, verification methods, and version
metadata. OAN does not need to embed every product-native file in the DID
Document. A Skill file, OpenAPI description, or MCP descriptor may live outside
OAN and be referenced by uniform resource identifier (URI), while Discovery
indexes the DID Document and
Root-verified metadata.

This resource model is intentionally narrower than a universal knowledge graph
and stronger than a simple endpoint registry. It does not try to describe every
domain ontology, pricing rule, runtime permission, or business workflow. It
standardizes the minimum trust envelope that different resources need before
they can be published across organizational boundaries: identity control,
resource type, endpoint or artifact reference, semantic description, lifecycle
state, governance scope, package provenance, and verification material.
Application communities can still define richer schemas above this envelope.

The model also avoids tying OAN to one artifact format. An Agent Service may
remain a live network service. A Skill may remain a portable package. An MCP
Server may keep its native protocol descriptor. A Tool/API may keep an OpenAPI
or domain-specific description. OAN does not erase those native forms. It gives
them a common identity, publication, and Discovery layer so relying parties can
ask the same pre-connection questions before they invoke different kinds of
resources.

This makes the resource model suitable for ecosystem growth. New product
communities can keep their native runtime or package format while exposing the
minimum OAN-facing facts needed for trust: who controls the resource identity,
what type of resource it is, which domain authorizes its exposure, which version
is current, and which Root-verified package should be used for Discovery and
verification. OAN can therefore serve as a shared trust envelope around diverse
Agent resources instead of forcing every resource into one marketplace schema
or one invocation protocol.

\subsection{Service Agent}

The Service Agent is a real business Agent that exposes callable capabilities.
It may provide domain knowledge, data access, workflow execution, tool access,
or task collaboration. In OAN, a Service Agent has an identity document and can
be registered, discovered, and invoked under trust controls.

The Service Agent should verify incoming trusted invocation envelopes. It should
check caller DID material, credentials, timestamp freshness, nonce uniqueness,
target DID binding, request body hash, and signature. These checks allow the
Service Agent to reject replay, substitution, wrong-target, and unauthenticated
requests before business logic runs.

\subsection{User Agent}

The User Agent queries Discovery, verifies signed discovery responses, selects
candidates, checks Root-verified package provenance, builds a signed invocation
request, and verifies signed Service Agent responses. The User Agent represents
the relying side of OAN. It turns Discovery results into verifiable interaction
decisions.

\begin{figure*}[!ht]
\centering
\includegraphics[width=0.95\linewidth]{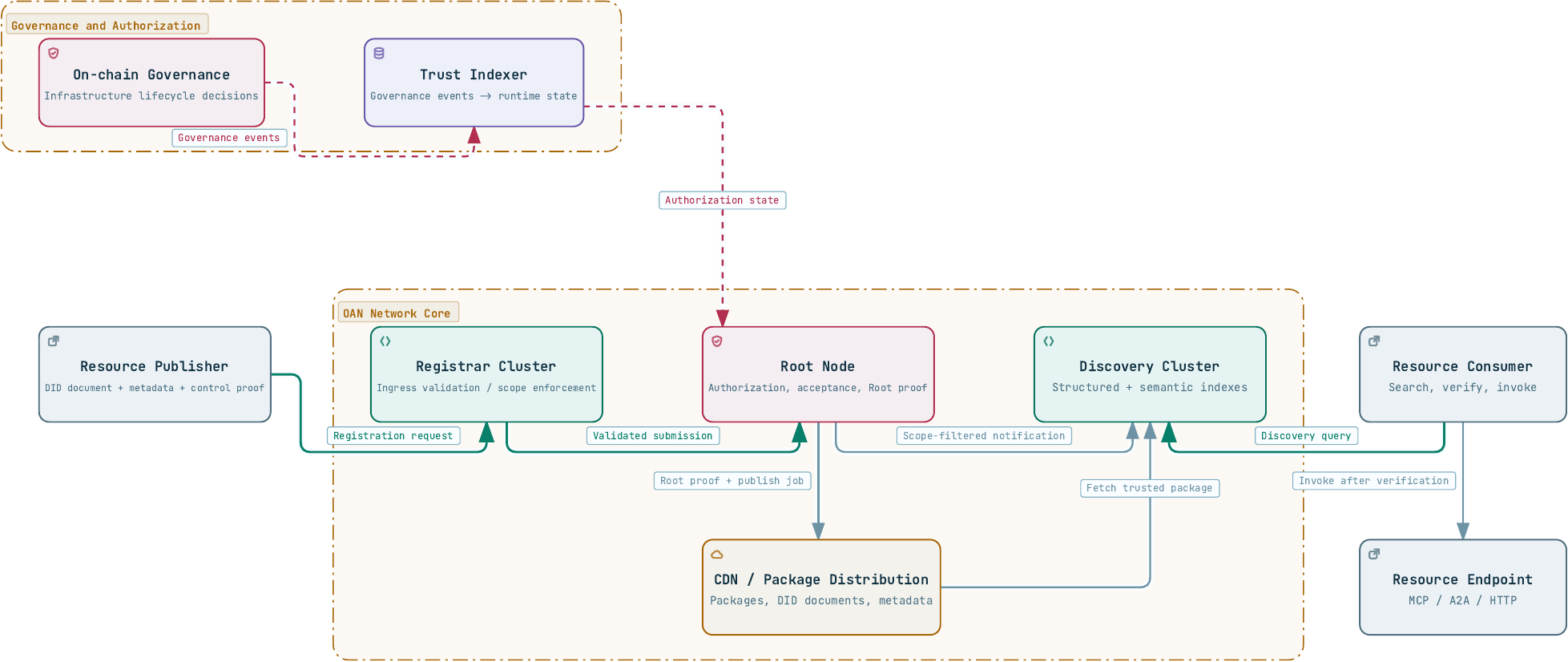}
\caption{OAN system roles and their trust relationships across governance, Root, Registrar, distribution, Discovery, and resource interaction.}
\label{fig:whitepaper-architecture}
\end{figure*}

The relationships among these roles are summarized in
Figure~\ref{fig:whitepaper-architecture}.

\section{OAN Trust Architecture}

OAN's trust architecture can be read as an evidence chain. The resource holder
first proves control over a \texttt{did:oan} identity. A Registrar adds
registration evidence after review and subject-control verification. Root
checks the submitted identity, Registrar authorization, lifecycle state, and
policy constraints before accepting a version. The distribution layer carries
Root-verified packages, while Discovery verifies those packages and applies
its own authorized-domain scope before returning signed results. Finally, a
relying User Agent and a Service Agent use signed invocation envelopes to bind
caller, target, request body, timestamp, nonce, and proof.

Trust verification requires evidence from multiple stages rather than a
directory entry alone. Each stage adds a different kind of
evidence: subject control, Registrar review, Root acceptance, package
integrity, Discovery authorization, response provenance, and invocation
freshness. If one stage is missing or stale, a relying party should be able to
detect that the candidate is not fully supported by the OAN trust path.

\subsection{Identity Representation}

An OAN resource identity representation is based on \texttt{did:oan} and is
compatible with DID-style concepts. It contains an identifier, verification
methods, product type, service endpoints or artifact references, semantic
description, capability metadata, credential references, OAN metadata, and
version-related information. It also carries typed \texttt{authorizedDomains}
when the represented subject or resource needs an explicit governance scope.
The representation supplies the material required for Root verification,
semantic Discovery, and runtime validation.

The \texttt{did:oan} identifier itself is routing-aware. Its method-specific
identifier contains a routing component and a uniqueness component. The routing
component helps Registrars, Root, and infrastructure nodes check that a DID is
being admitted through the expected authority path, but it is not a substitute
for proof, VC status, or governance state. This keeps the identifier useful for
operational routing while preserving the security rule that trust must come
from signatures, credentials, Root acceptance, and current authorization state.

\texttt{did:oan} is deliberately resource-oriented. It can describe Agent
Services, Skills, MCP Servers, and Tool/API resources without changing the
basic trust path. The holder behind the resource still controls private keys
and credentials. For callable resources, the holder can use those keys during
mutual verification before invocation. For downloadable or descriptor-like
resources, the DID Document can provide semantic discovery fields and external
artifact references while keeping artifact distribution outside the core OAN
protocol.

The identity representation supports three needs at once. First, it provides
human-readable and machine-readable metadata for discovery and integration.
Second, it provides cryptographic verification material. Third, it gives Root
and Discovery enough information to apply lifecycle and authorization checks.
Authorization domains and capability tags serve distinct purposes. A tag may
help describe or rank a resource, while
\texttt{authorizedDomains} is the field that Root, Registrar, Discovery, and
governance tooling use for scope enforcement.

The same identity model also supports external identifier compatibility. A DID
Document may declare external identifiers from existing platforms or industry
identifier systems. The external identifier's native \texttt{id} remains a
metadata value under its own resolution context; it does not replace the OAN
DID and does not grant controller authority. This allows OAN to connect
existing resource catalogs to a verifiable OAN trust path without forcing every
external platform to change its native identifier scheme.

Controller relationships are similarly explicit. A self-controlled resource
uses its own OAN Identity to prove control. A dedicated controller is itself an
OAN-registered identity that may authorize registrations for other resources.
When a resource points to an external controller, the controller value alone is
only a declaration; the registration path must verify a controller
authorization proof over the resource, controller, Registrar, document hashes,
purpose, validity window, and nonce. This prevents a publisher from claiming an
arbitrary controller DID without showing control of that controller's key.
For an update, the authorizing identity is the controller in the current
Root-accepted version, not the controller proposed by the new document. Thus,
if controller $A$ transfers a resource to $B$, $A$ authorizes the transition
and $B$ authorizes later versions. Root advances the resource version and
controller state atomically, preventing a superseded controller from updating
the resource after transfer.

\subsection{Registration Credential}

The registration credential links Registrar review and resource identity. It
binds the Registrar DID, resource DID, credential type, subject-control
evidence, validity, product type, and proof. Root verifies this credential
before accepting a submitted identity representation. This creates a separation
between review workflow and Root acceptance while preserving verifiability.
The credential attests the onboarding facts defined by its profile. It does not
by itself prove a legal-person or enterprise identity unless the Registrar's
policy and credential claims explicitly provide that assurance.

\subsection{Root Acceptance}

Root acceptance is the point where a resource identity version becomes
authoritative within a trust domain. Root verifies the submitted representation,
credential, signed upstream request, Registrar authorization, DID structure,
hashes, capability metadata, policy constraints, and, for an update, continuity
from the current controller. If accepted, Root atomically records the version
and current controller state and publishes verifiable evidence. If rejected,
the reason should be auditable.

\subsection{Root-Verified Package}

A Root-verified package is the distribution artifact consumed by CDN and
Discovery. It contains the identity document or references, hashes, metadata,
status, publication cursor, Root proof, and capability information. Discovery
does not trust a package because it came from a CDN; it verifies the package
against Root proof and bulletin facts.

\subsection{Signed Discovery Response}

Discovery responses are signed. A relying User Agent can verify that a response
came from the Discovery node that produced it, that the response is fresh, and
that returned candidates carry provenance sufficient for later validation.

\subsection{Identity Directory for Application Login}

Some applications need a lightweight way to let a controller or resource use an
OAN Identity for login. OAN supports this by allowing applications to query a
Root-accepted public-key projection for accepted controller and resource
identities. This projection is not a new authorization authority. It helps an
application find the public key associated with an accepted DID, after which
the application must still verify a fresh challenge signature from the user's
OAN Identity. The directory does not replace DID resolution, VC status checks,
controller authorization proof, or Root package verification.

\subsection{Trusted Invocation}

Trusted invocation extends OAN from discovery into the first step of
interaction. A User Agent signs a request envelope that binds caller DID, target
DID, timestamp, nonce, body hash, method, path, and proof. The Service Agent
verifies the envelope before executing business logic. OAN does not dictate the
complete business protocol after this point, but it establishes a trustworthy
entry condition.

\begin{figure*}[!ht]
\centering
\includegraphics[width=0.95\linewidth]{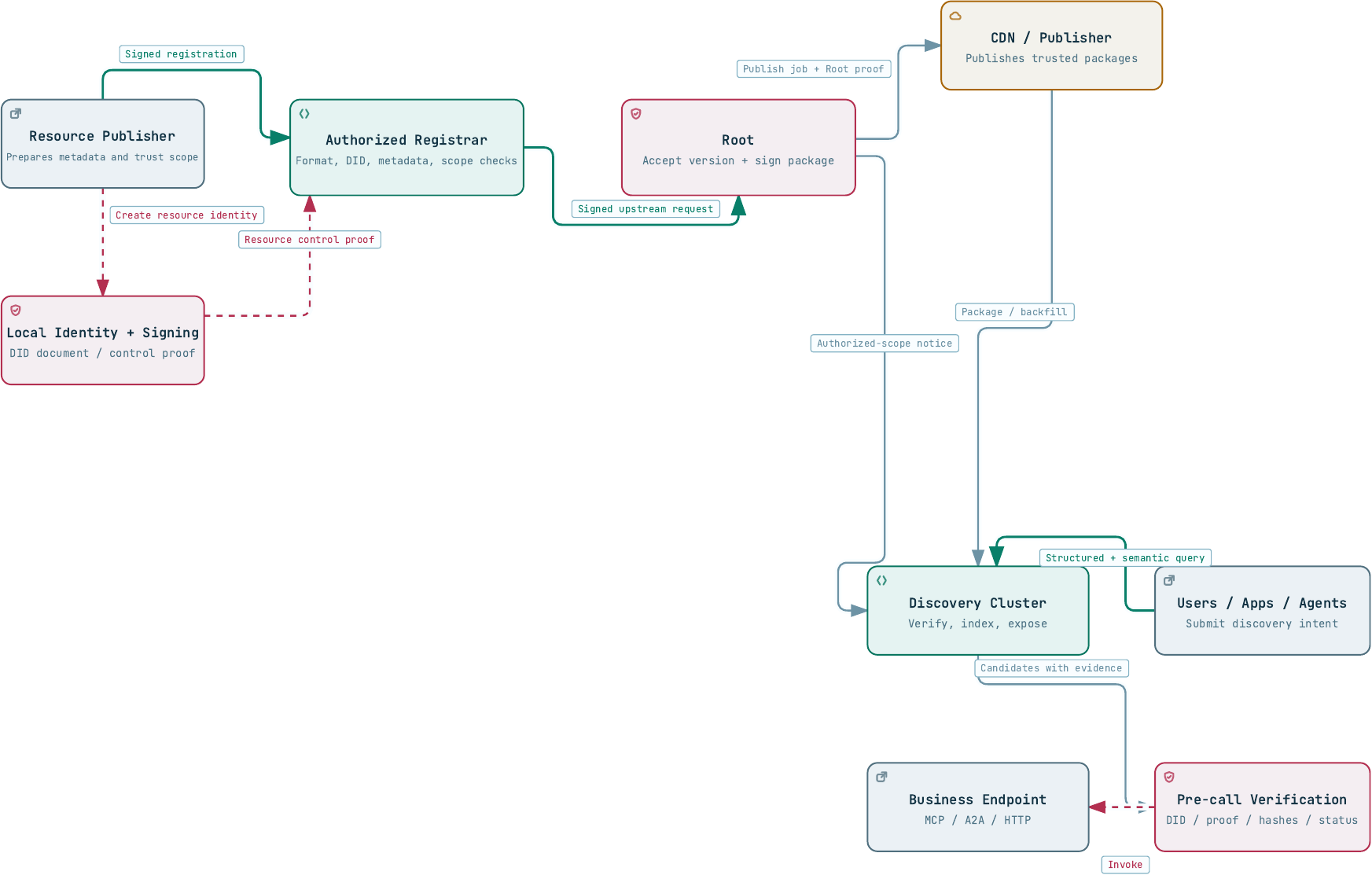}
\caption{The OAN trust path from resource registration through Root-verified publication and authorized Discovery to verified invocation.}
\label{fig:whitepaper-trust-flow}
\end{figure*}

Figure~\ref{fig:whitepaper-trust-flow} follows this evidence path from
identity preparation to the first verified invocation.

\section{OAN as an Adoptable Identity Infrastructure}

OAN is designed to be an identity infrastructure that heterogeneous Agents can
adopt without replacing the interaction protocols already used by their
applications. Its governed resource identity, trusted Discovery, and signed
invocation establish a common trust context for resources that may otherwise
belong to different Agent runtimes, tool systems, and service deployments. The
same resource-oriented model can therefore be used for an Agent Service, an
MCP Server, a Skill, or a Tool/API while leaving the resource's native business
semantics in place. Adoption depth is determined by the verifiable and enforced
path that an implementation actually uses; the mere presence of a DID, SDK, or
deployment component does not by itself establish L1 or L2 assurance.

Adoption separates public evidence from local identity custody. A relying party
can obtain a DID Document, registration credential, Root-verified package, and
Discovery result as verifiable public material. The corresponding OAN Identity
and its private signing key remain under the control of the user or resource
operator. This separation allows an Agent to use OAN identities without
turning a public registry into a centralized key holder, and it allows a
controller to manage a resource relationship without becoming the resource's
business signer.

\subsection{Progressive Integration Depths}

OAN supports progressive integration depths so that adoption can begin with
identity publication and continue toward protected protocol execution.

At \textbf{L0}, an Agent or resource is registered and discoverable. Its OAN
DID, service endpoint or artifact reference, resource type, and public metadata
are bound to the governed registration and Discovery path. L0 gives an application a trusted
candidate and a way to resolve its public identity; it does not by itself
prove that a live business request was signed by the resource.

At \textbf{L1}, the adopting application verifies the identity evidence before
interaction. It checks the DID Document, registration credential,
Root-verified package, Discovery provenance, current version, and the binding
between the selected endpoint and resource identity. Existing channel
mechanisms such as Transport Layer Security (TLS) or OAuth may remain in use,
while OAN supplies the
identity and governance context that the channel alone does not provide.

At \textbf{L2}, both communicating resources use their own OAN identities in
the selected protocol path. The request, response, and applicable events are
bound to the sender, target, operation, audience, freshness, and replay
controls, and authorization is checked before the business handler accepts the
operation. L2 therefore carries OAN identity into an actual interaction rather
than stopping at discovery or a preliminary handshake.

At \textbf{L3}, an operator extends L2 with deployment and business controls
such as bounded delegation, multi-tenant policy, protected private-network
boundaries, remote or hardware-backed key custody, confidential payload
handling, and stronger runtime or instance assurances. L3 is an extension of
the identity and message-binding foundation, not a replacement for the native
Agent protocol.

\subsection{Integration Paths for Existing Agents}

\begin{figure*}[!ht]
\centering
\includegraphics[width=0.8\linewidth]{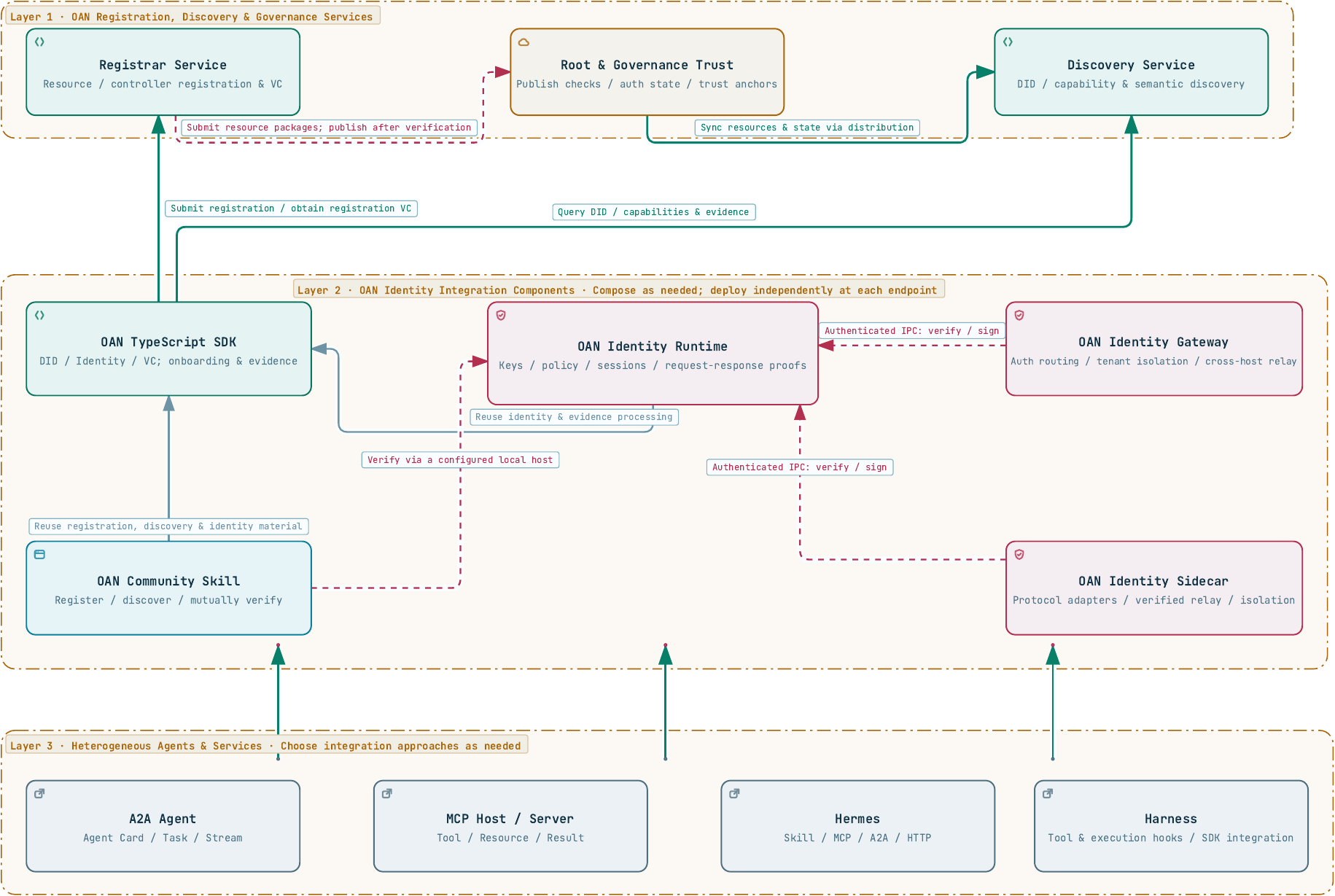}
\caption{OAN identity integration architecture. Composable adoption interfaces
connect heterogeneous Agents and services to OAN infrastructure while retaining
their native protocols and endpoint identity boundaries.}
\label{fig:wp-identity-integration}
\end{figure*}

Figure~\ref{fig:wp-identity-integration} shows how the SDK, Skill, Runtime,
Sidecar, and Gateway connect heterogeneous Agents and services to OAN's
registration, discovery, and governance services. Integrators compose these
interfaces according to the identity capabilities required at each endpoint.

The Community Skill provides a low-friction path for an assistant that can
execute skills or local tools. It can guide registration, Discovery, material
handoff, and a two-sided identity challenge by invoking local identity
facilities. A Skill can make a real identity check understandable to a user,
but it does not enforce protection for every subsequent business message;
that property depends on the selected runtime or protocol boundary.

The Identity Runtime is the local identity and key-use layer. It maintains a
catalog of identities, invokes a controlled key provider, evaluates policy,
keeps session and replay state, and verifies OAN evidence. It gives a new Agent
implementation a direct integration path and gives a Skill or adapter a common
local service without exposing private keys to prompts, remote services, or
business handlers.

A Sidecar offers a lower-change path for an existing Agent or service. It can
place a verifiable ingress and egress boundary around the native endpoint,
obtain signing and verification decisions from the local Runtime, and pass
only verified caller context to the backend. A Gateway extends this model
across private networks and organizational boundaries. It can preserve the
original caller, proxy, and target resource identities, apply tenant and
delegation policy, and keep channel authentication separate from business
identity.

These paths compose rather than compete. A Skill may use the Runtime for local
identity operations; a native MCP or A2A implementation may use the Runtime
directly; a legacy service may use a Sidecar; and a multi-tenant deployment may
place a Gateway in front of one or more Sidecars. In every case, the resource
identity used for a business message remains distinct from any controller,
proxy, or transport identity.

\subsection{MCP and A2A Adoption}

MCP and A2A provide the principal protocol surfaces for carrying OAN identity
into Agent interaction. For MCP, an implementation can bind the verified OAN
resource identity to the server endpoint, tool selection, invocation
arguments, result or error, and stream or cancellation context. MCP continues
to define tool and resource semantics, while OAN provides the governed peer
identity and message trust context.

For A2A, an implementation can bind the OAN resource identity to the Agent
Card, task exchange, artifacts, events, callbacks, and final task state. A2A
continues to define task and collaboration semantics, while OAN supplies
identity provenance, endpoint binding, and the signatures needed to associate
messages with the intended resource.

This separation lets heterogeneous Agents adopt OAN at the depth appropriate
to their deployment. An application may begin with L0 Discovery, add L1
evidence checks, use a Runtime for mutual identity verification, and introduce
Sidecar or Gateway enforcement when L2 or L3 controls are required. OAN thus
provides a common identity foundation while preserving the diversity of Agent
protocol implementations.

\section{Authorization Domains and Controlled Discovery}

Authorization domains are the governance bridge between identity metadata and
discovery exposure. A resource may describe many capabilities, but it should
only be accepted by a Registrar whose authorized scope covers the resource's
\texttt{authorizedDomains}; after Root acceptance, it should only be delivered
to Discovery nodes whose authorized scope covers those same domains. This is
essential for multi-operator deployment. It allows a trust domain to operate
domain-specific Registrars and Discovery services without giving each service
global exposure power.

Authorization-domain governance separates three scope-related layers from
local search refinements:

\begin{itemize}[leftmargin=*]
  \item \textbf{canonical domains}: a governed domain taxonomy that provides
  stable top-level and domain-specific categories;
  \item \textbf{infrastructure authorization domains}: the domain set assigned
  to each Registrar and Discovery node through governance and Root-issued
  infrastructure authorization VCs;
  \item \textbf{resource authorization domains}: the \texttt{authorizedDomains}
  field bound into the resource DID Document, metadata, Root package, and
  Registrar-issued resource credential;
  \item \textbf{local refinement tags}: custom tags, descriptions, indexes, or
  ranking signals used after authorization checks.
\end{itemize}

The key principle is that custom tags cannot expand governance authority. A
Registrar may suggest tags to improve description quality, and a Discovery node
may use custom metadata to improve search quality inside an authorized domain,
but neither tags nor semantic search may create authorization outside
\texttt{authorizedDomains}. The special value \texttt{["*"]} denotes explicit
all-domain authorization, while an empty list denotes no domain authorization.
A parent domain covers its descendants by hierarchy, so excluding a child
domain also requires excluding any parent that would cover it.

This separation is important for semantic search. Embeddings, small-language
model indexes, domain ontologies, and hybrid matching algorithms can improve
how a user requirement is mapped to resource candidates, as explored by
GRAIL~\cite{xu2026grail}. However, they are ranking and retrieval mechanisms
inside an already authorized result space. They should not expand a Discovery
node's governed exposure scope.

\section{Relationship with MCP, A2A, and ANP}

OAN is most useful when understood as a layer around existing protocols rather
than a competitor to them.

\subsection{MCP}

The Model Context Protocol focuses on connecting AI applications with tools and
data sources~\cite{mcp_spec}. It gives clients a structured way to discover
and call tools provided by MCP servers. OAN can complement MCP by providing
identity and trust context before a tool server or client is used. For example,
an Agent can use OAN to verify that an MCP server endpoint belongs to a
Root-accepted identity, that the endpoint is current, and that the peer
identity has appropriate credentials. MCP remains responsible for tool
protocol details.

\subsection{A2A}

Agent-to-Agent protocols focus on task exchange, capability description,
collaboration state, and message flow between Agents~\cite{a2a_spec}. OAN can provide trusted
identity onboarding, discovery, and pre-connection validation for
A2A-compatible Agents. In a layered deployment, an Agent might first use OAN to
verify a discovered peer and then open an A2A session for task negotiation.

\subsection{ANP and Agent DID Methods}

ANP-like work and Agent-oriented DID methods explore identity, networking, and
interoperability for Agents~\cite{did_wba_spec}. OAN is aligned with
DID/VC-style identity and can interoperate with Agent DID methods. Its
distinctive focus is the complete governed lifecycle: Registrar authorization,
Root acceptance, package publication, capability-scoped Discovery, and signed
invocation.

\subsection{Layering Summary}

Table~\ref{tab:protocol-relation} summarizes the relationship: OAN answers whether a resource
identity is admissible, current, and discoverable under policy; MCP answers how
tools are exposed and called; A2A answers how Agents exchange tasks; ANP and
DID methods can provide identity and networking conventions; domain protocols
answer how specific industries structure their data and workflows.

\begin{table}[t]
\centering
\caption{OAN Compared with Adjacent Protocol Layers}
\label{tab:protocol-relation}
\resizebox{\columnwidth}{!}{%
\begin{tabular}{@{}lll@{}}
\toprule
Layer & Main concern & OAN relationship \\
\midrule
MCP & tool and data access & adds identity trust \\
A2A & Agent task exchange & adds discovery trust \\
ANP & Agent networking & compatible identity layer \\
DID/VC & identifier and credential substrate & reused by OAN \\
Domain APIs & business semantics & can run above OAN \\
\bottomrule
\end{tabular}
}
\end{table}

This separation defines OAN's protocol compatibility boundary. MCP, A2A, ANP,
representational state transfer (REST) APIs and domain-specific protocols can
all remain responsible for their
own wire behavior. OAN adds a shared trust context through
a resolvable resource DID, Root-verified package evidence, Discovery proof, and
credential or signature checks appropriate to the resource type. This keeps
OAN broad enough to cover many products without turning it into a universal
runtime.

\subsection{Relationship with Registries, Directories, and Schemas}

The Agent ecosystem is also developing registries, directories, and schema
frameworks that are closer to OAN's problem space than invocation protocols.
OAN should be understood as complementary to these efforts, but with a
different primary object and trust boundary.

MCP registry work is primarily concerned with making MCP servers and their
metadata discoverable~\cite{mcp_registry}. OAN can represent an MCP Server as a
resource type and bind its endpoint to a \texttt{did:oan} identity, Root
acceptance, and verifiable package. In this sense, an MCP registry can be
carried by OAN, while OAN's trust path is broader than MCP server lookup.

Skill marketplaces focus on packaging, browsing, and distributing Skills. OAN
does not need to replace such marketplaces. Its contribution is to make a Skill
a first-class governed resource: a Skill can have a DID, semantic metadata,
version state, package proof, and discovery evidence even if the Skill artifact
itself is hosted by another marketplace or repository.

Agent schema frameworks aim at common languages for Agentic records,
capabilities, Skills, and relationships~\cite{agntcy_oasf}. OAN's
authorization-domain taxonomy and semantic resource fields solve a neighboring
problem: how accepted resources are classified and discovered under governance.
A future OAN deployment can embed or reference compatible schema records inside
a ResourcePackage instead of requiring a separate OAN-only schema vocabulary.

The Agent Directory Service (ADS) is closer to an Agent directory and discovery fabric, including directory,
metadata, routing, and content-distribution concerns~\cite{agntcy_ads_overview}.
The difference is emphasis. ADS is directory-first: it starts from scalable
discovery of Agent metadata. OAN is trust-first: it starts from governed
registration, Root acceptance, verifiable distribution, authorization-aware
Discovery, and signed use of results. The two approaches can converge when ADS
or external schema records are treated as resource metadata inside OAN's trust
envelope.

The Agent Name Service (ANS) and related Agent naming systems focus on name assignment and name-based
discovery for Agents~\cite{huang2025ans,cui2025agentdns}. OAN is DID-first and
resource-first. It can interoperate with names as aliases or service metadata,
but its core object is the trusted resource identity and ResourcePackage rather
than the name alone. Table~\ref{tab:adjacent-positioning} contrasts these
primary objects and their relationship to OAN.

\begin{table}[t]
\centering
\caption{Positioning of OAN Against Adjacent Systems}
\label{tab:adjacent-positioning}
\scriptsize
\begin{tabular}{@{}p{0.25\columnwidth}p{0.30\columnwidth}p{0.35\columnwidth}@{}}
\toprule
System family & Primary object & OAN relationship \\
\midrule
MCP registry & MCP server/tool endpoint & resource type and trust wrapper \\
Skill marketplace & Skill artifact & governed Skill identity \\
Agent schema & Agentic schema & compatible metadata schema \\
ADS & Agent metadata directory & trust-first registry and discovery \\
ANS & Agent name & complementary alias/name layer \\
Harness/runtime & task execution & outside OAN data plane \\
\bottomrule
\end{tabular}
\end{table}

\begin{figure*}[!ht]
\centering
\includegraphics[width=0.8\linewidth]{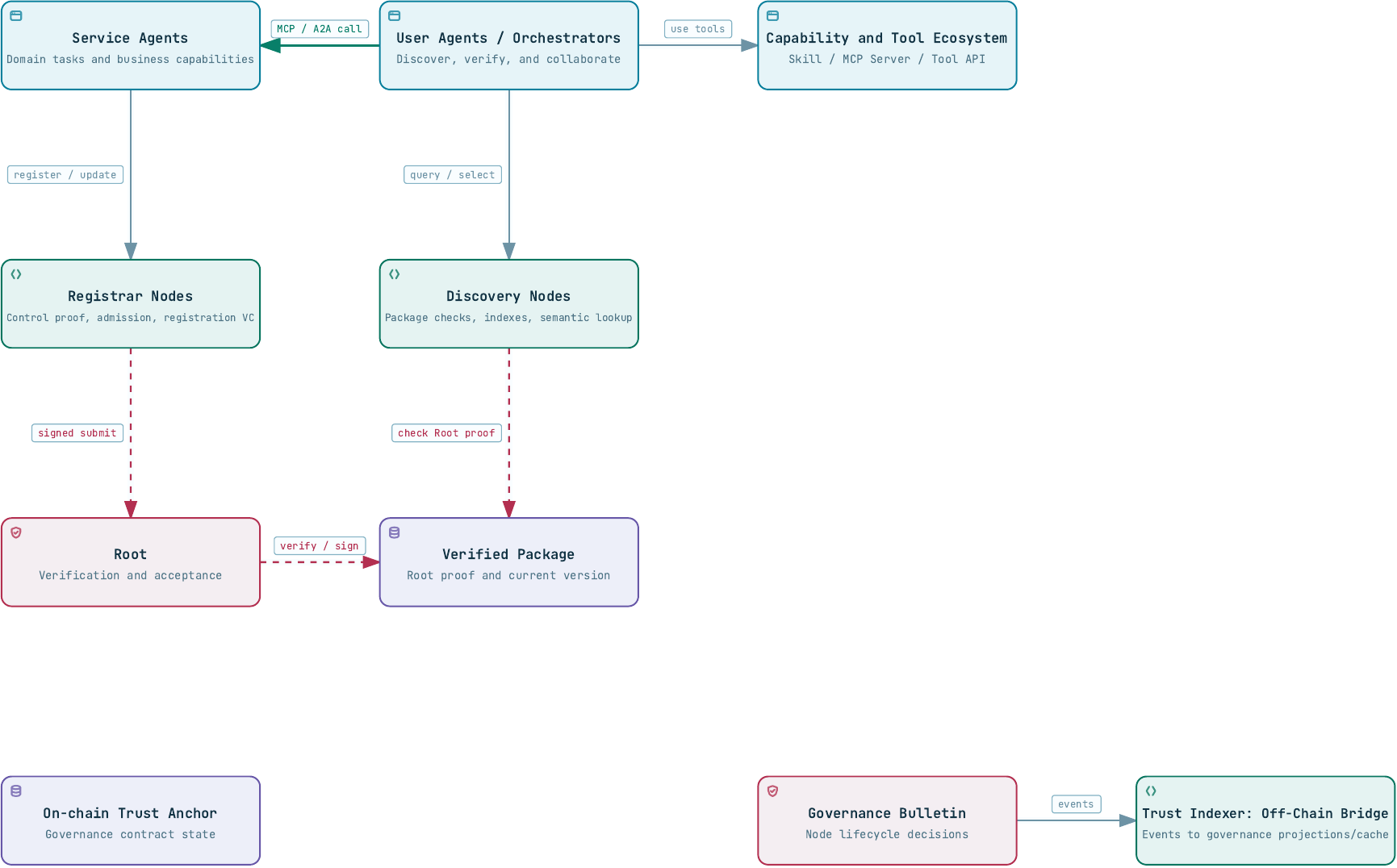}
\caption{OAN supplies identity and trust beneath Agent interaction protocols while applications retain their own task and workflow semantics.}
\label{fig:whitepaper-layering}
\end{figure*}

Figure~\ref{fig:whitepaper-layering} places OAN's trust functions beneath the
interaction protocols and above the cryptographic primitives.

This layering is the core positioning of OAN. It lets a partner keep an MCP
server, an A2A-capable Agent, an ANP-like route structure, or a domain API while
adding a common trust path before discovery and invocation. OAN should therefore
be evaluated not as a rival to those protocols, but as the layer that can make
cross-protocol Agent discovery and first contact governable.

\section{Operational Deployment Patterns}

\subsection{Single-Operator Trial Network}

The simplest deployment places Root, Registrar, Discovery, CDN, Service Agent,
and User Agent under one operator. This pattern is useful for development,
testing, demonstration, and early governance experiments. It validates lifecycle
correctness without introducing cross-organization operations too early.

\subsection{Federated Infrastructure Governance}

In a multi-organization deployment, governance should not be reduced to the
private decision of one Root operator. OAN therefore separates committee-level
infrastructure lifecycle decisions from Root-side protocol enforcement. A
governance committee can authorize, suspend, recover, or revoke Registrar,
Discovery, and future VC issuer nodes through threshold-governed actions
published by the governance layer. Root observes these decisions and issues or
refuses infrastructure authorization VCs accordingly.

This pattern changes the system from a single-operator trust path into a
federated trust-governance architecture. Root still anchors accepted resource versions
and signs packages, but it does not need to be the sole institution that
controls which infrastructure participants may exist. That distinction is
important for standards-facing deployment, because partner organizations can
participate in governance without forcing ordinary Agents or resource
providers to understand governance operation code.

\subsection{Multi-Registrar Onboarding}

As the ecosystem grows, multiple Registrars can onboard Agents for different
communities or domains. Root remains the final resource-acceptance and
package-signing authority inside the trust domain, while Registrar lifecycle
authorization is governed above Root. This pattern allows domain operators to
perform onboarding work while keeping a consistent trust record.
Registrar authorization is domain-scoped. A Registrar may accept a resource
only when the resource's \texttt{authorizedDomains} are covered by the
Registrar's own authorized domain set. This prevents a domain-specific
Registrar from becoming a global admission gateway. When a resource does not
explicitly provide domains, an implementation may reject the submission,
default it to the Registrar's approved scope, or derive a reviewed subset,
depending on local onboarding policy; in every case the final domains should be
explicit in the accepted representation.

\subsection{Domain-Specific Discovery}

Different Discovery nodes can serve different authorized domains. For example,
one Discovery node may focus on industrial Agents, another on research Agents,
and another on public-service Agents. Each Discovery node receives only the
authorization it needs.
Root targets Discovery notification by intersecting resource
\texttt{authorizedDomains} with each Discovery node's authorized domains.
Discovery then verifies the Root package and refuses to index packages outside
its scope. Query matching therefore runs after governance-bound exposure, not
before it.

\subsection{Partner Integration}

A partner organization can keep its existing Agent runtime and protocol stack
while adding OAN registration, discovery, and verification. It may expose an
OAN-compatible identity document, register through a Registrar, synchronize
with an authorized Discovery node, and add verification before MCP or A2A
interaction.

\subsection{Production Deployment}

Production deployment combines the protocol trust path with database
resilience, key management, observability, rate limiting, durable replay
protection, backup, monitoring, deployment automation, and conformance testing.
Deployment profiles configure these mechanisms according to operational scale,
security requirements, and availability targets.

\section{Governance Model}

OAN governance is divided between committee-level infrastructure decisions,
Root-side policy enforcement, and protocol evidence. Policy decisions include
who may operate a Registrar, which domains a Registrar may admit, which domains
a Discovery node may expose, how authorization domains are maintained, what
evidence is required for registration, and how revocation is handled. Protocol
evidence turns these decisions into signed, verifiable facts.

\subsection{Infrastructure Authorization}

Infrastructure authorization has two linked layers. The on-chain governance
layer records lifecycle state for Registrar, Discovery, and future third-party
VC issuer nodes. Root then issues infrastructure authorization VCs to nodes
whose governance state is active and whose DID-control and operational
materials pass Root verification.

Neither layer is sufficient alone. A Registrar with a valid Root VC should not
be accepted after its governance state becomes inactive. A Discovery node whose
governance state is active should not be treated as protocol-authorized before
Root has issued the corresponding VC. Effective authorization is therefore:

\[
  \textit{active governance state} \land \textit{valid Root-issued VC}.
\]

This distinction lets the governance layer provide auditable lifecycle changes
while Root remains the issuer and verifier of protocol credentials. In this
sense, OAN is closer to a federated trust-governance infrastructure than to a
plain Agent registry. The governance layer answers who is allowed to operate
infrastructure; Root enforces that answer and signs the protocol artifacts used
by services; Discovery exposes only Root-accepted and policy-admissible
resource identities.

\subsection{On-Chain Governance Layer}

For infrastructure participants, OAN can publish lifecycle authorization
changes through an on-chain governance layer. The layer records and emits
events for Registrar, Discovery, and future third-party VC issuer lifecycle
changes, including authorization, suspension, recovery, revocation, and
Registrar/Discovery authorized-domain updates. Infrastructure services consume the event stream
through an indexer or local cache, rather than asking Root for current status on
every request.

This layer is not a storage layer for resource identities. It does not store,
parse, validate, or issue DID Documents, Verifiable Credentials, metadata, or
capability descriptions. These remain OAN protocol and service
responsibilities. The governance layer may carry external references such as
subject identifiers and hashes of review material, but its core function is
participant lifecycle management.

This boundary is deliberate. Governance events are low-frequency and
high-value: node authorization, suspension, recovery, revocation, and
authorized-domain changes should be auditable and hard to rewrite. Resource
registration, package synchronization, semantic indexing, and Discovery queries
are higher-frequency runtime activities and should remain off chain. OAN uses
the chain as an authorization and audit substrate, while Root, Registrar,
Discovery, and indexer services translate those facts into efficient runtime
behavior.

The governance operation model keeps proposal creation, review, voting, and
lifecycle actions in designated governance clients or processes, separate from
ordinary Registrar, Discovery, and VC issuer service runtimes. The intended
direction is federated committee operation with threshold approval, so the
system can move from initial stewardship toward broader institutional
participation. Registrar, Discovery, and VC issuer service nodes are event
consumers by default. As the ecosystem matures, governance operation clients can
be opened to more participating organizations without changing ordinary
resource-provider or service-node integration.

\subsection{Resource Admission}

Resource admission involves both operator workflow and protocol checks. A
Registrar may help review metadata and verify subject control. Root then
verifies the complete representation and supporting evidence. Admission should
not depend only on a website account, email address, or human-readable
description. The same admission path can handle Agent Service, Skill, MCP
Server, and Tool/API products because their OAN-facing representation is a
\texttt{did:oan} DID Document plus verifiable registration evidence.

\subsection{Versioning and Revocation}

Resources change. Endpoints move, keys rotate, capability descriptions evolve,
Skills gain new releases, APIs add or remove operations, and operators may be
revoked. OAN treats identity as versioned state. A current version should
supersede older accepted versions, and revoked or inactive versions should not
be discoverable after synchronization.

\subsection{Auditability}

Operators should be able to answer who submitted an identity, which Registrar
issued the credential, which Root decision accepted or rejected it, which
package was published, which Discovery node indexed it, and why a request was
accepted or rejected. This audit path is central to cross-organization trust.

\section{Security and Risk Posture}

OAN applies zero-trust and workload-identity principles to Agent
interconnection~\cite{rose2020zerotrust,spiffe_spire}. It establishes verifiable
identity, governance, Discovery, and pre-connection evidence on which
applications can build business authorization and safety policies. These
infrastructure guarantees complement application-specific assessments of
correctness, truthfulness, safety, and service quality.

\subsection{Threats Addressed}

The architecture is designed to reduce several risks:

\begin{itemize}[leftmargin=*]
  \item identity imitation through unsigned or unaccepted metadata;
  \item controller impersonation through unsupported controller strings;
  \item stale endpoint use after identity update or revocation;
  \item unauthorized Registrar submission;
  \item Discovery exposure outside authorized domains;
  \item tampering with distributed packages;
  \item replay or wrong-target invocation requests;
  \item blind trust in directory output without provenance verification.
\end{itemize}

\subsection{Threats Not Fully Addressed}

Prompt injection, model hallucination, malicious tool behavior, data leakage
inside business logic, semantic misrepresentation of capabilities, and economic
reputation require application, model, policy, and operational controls. OAN
provides verifiable identities and evidence bindings through which those
controls can be associated with specific resources and responsible parties.
A compromised Root signing boundary or captured governance process can also
produce apparently valid trust-domain evidence; key custody, threshold
governance, audit, and incident response reduce this risk but do not remove the
Root trust assumption within the current single-domain model.

\section{Prototype and Evaluation Status}

The current performance baseline follows resources from Registrar submission
through Root acceptance, CDN publication, and authorized Discovery visibility.
It crosses Agent-only and mixed-resource batches with two logical topologies:
one Registrar and one Discovery node, or three Registrars and two Discovery
nodes. Both topologies include one Root, one CDN node, and one publication
worker.

\paragraph{Experimental setting and measurement}
The workloads ran serially on a local Windows host using release builds,
PostgreSQL~17, NATS JetStream, four client worker processes, and 48 total
in-flight registrations. Each scenario and scale used fresh isolated state;
resource construction, compilation, and service startup were outside the
measurement window. Root, CDN, and each Discovery count were sampled once per
second by a read-only observer. The multi-node configuration therefore tests
role composition and independent Discovery convergence on one host, not
geographic scale-out.

Table~\ref{tab:evaluation-coverage} lists the four workloads and their actual
batch sizes. Each workload was repeated five times at three target scales, giving 60
isolated runs. The multi-node Agent workload distributes equal counts across
three Registrars and therefore uses actual totals of 498, 1998, and 4998; the
other workloads use 500, 2000, and 5000 resources. Registration throughput is
based on client completion time. End-to-end throughput uses the first sampled
time at which Root, CDN, and all Discovery targets reach the expected deduplicated
count. It does not include per-resource capability queries or Agent task
execution.

\begin{table}[!ht]
\centering
\caption{Performance Workloads and Batch Scales}
\label{tab:evaluation-coverage}
\small
\begin{tabular}{@{}lll@{}}
\toprule
Workload & Registrar/Discovery & Actual resources \\
\midrule
Single / agents & 1 / 1 & 500, 2000, 5000 \\
Single / mixed & 1 / 1 & 500, 2000, 5000 \\
Logical multi / agents & 3 / 2 & 498, 1998, 4998 \\
Logical multi / mixed & 3 / 2 & 500, 2000, 5000 \\
\bottomrule
\end{tabular}
\end{table}

\begin{figure}[!ht]
\centering
\includegraphics[width=0.47\textwidth]{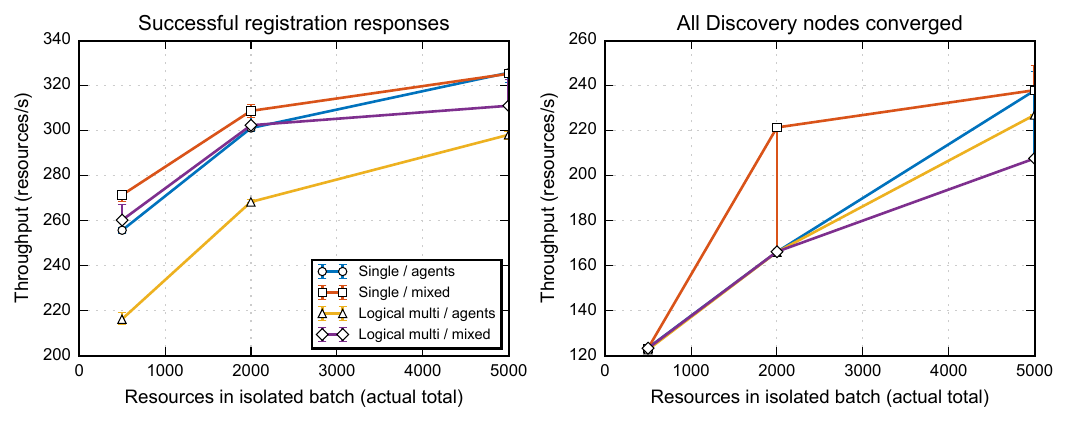}
\caption{Median Root-acceptance and structured end-to-end throughput at the
largest scale. Error bars show the observed range across five isolated runs.}
\label{fig:wp-pipeline-throughput}
\end{figure}

\paragraph{Throughput and progression at the largest scale}
At approximately 5000 resources, median end-to-end completion times are
21.021, 21.011, 22.021, and 24.077~s for single/Agent, single/mixed,
logical-multi/Agent, and logical-multi/mixed workloads. The corresponding
median end-to-end rates are 237.86, 237.97, 226.97, and 207.67 resources/s.
Figure~\ref{fig:wp-pipeline-throughput} reports both Root acceptance and
end-to-end throughput; medians are calculated from five independent runs.

\begin{figure}[!ht]
\centering
\includegraphics[width=0.47\textwidth]{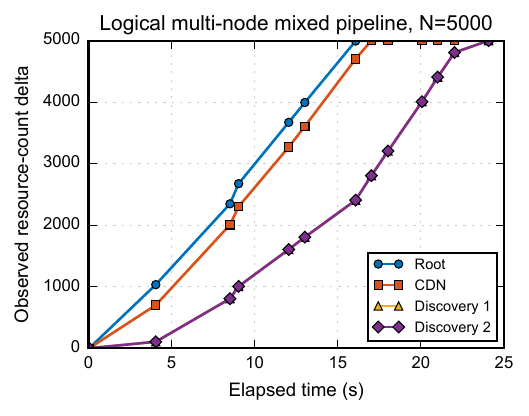}
\caption{Root, CDN, and per-Discovery progress for representative largest-scale
runs. Sampling interval is one second; completion times are therefore sampled
upper bounds rather than server-internal event timestamps.}
\label{fig:wp-pipeline-progress}
\end{figure}

The one-second samples also separate the three asynchronous stages. At the
largest scale, the median maximum Root--CDN count gaps are 345, 230, 191, and
650 resources across the four workloads; the corresponding CDN--slowest
Discovery gaps are 1984, 1889, 2013, and 2296. These are sampled count
differences, not exact internal queue lengths. Figure~\ref{fig:wp-pipeline-progress}
shows the representative progression data without combining the two Discovery
targets in the logical multi-node workloads.

\paragraph{Completeness and visibility}
All 60 runs reached the expected deduplicated Root, CDN, and Discovery counts,
with a median sampling-error count of zero in every scenario--scale group.
Mixed resources follow the same lifecycle as Agent-only resources, and each
Discovery target independently reaches the full expected total. The experiment
measures structured registration, publication, and visibility. Semantic-index
convergence and retrieval quality require a separate pgvector-enabled
evaluation and are not inferred from these results.

\subsection{Cross-Host Deployment Validation}

A separate experiment exercised the propagation path across three physical
hosts. One laptop ran the registration driver and one Registrar; a second
laptop ran another Registrar and one Discovery node; and a cloud host ran the
Root, a third Registrar, a second Discovery node, the CDN, publication worker,
and NATS. PostgreSQL remained local to each node. With 16 in-flight
registrations, four batch sizes (30, 300, 1200, and 2000) were each repeated
five times from fresh identities, databases, and namespaces. This experiment
predates and is methodologically separate from the 60-run single-host matrix
above.

\begin{figure}[!ht]
\centering
\includegraphics[width=0.47\textwidth]{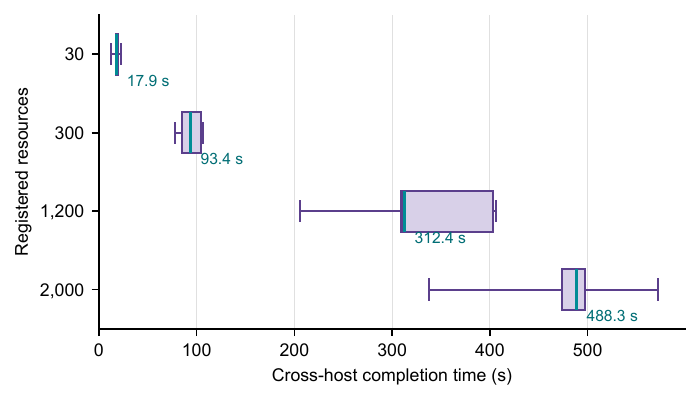}
\caption{Cross-host completion-time distributions for a fixed three-host
deployment. Each box descriptively summarizes five isolated runs and uses the
observed minimum and maximum as whiskers; completion requires both Discovery
nodes to reach the expected resource count.}
\label{fig:wp-cross-host-completion}
\end{figure}

Both Discovery nodes reached the expected count in all 20 runs. Median
cross-host completion times were 17.945, 93.411, 312.387, and 488.306~s as the
batch size increased. Figure~\ref{fig:wp-cross-host-completion} summarizes the
five completion observations at each scale. Completion includes registration
and downstream publication, retrieval, validation, and visibility; it does
not isolate network-transfer time or establish hardware-matched scale-out
speedup. Because the implementation version, concurrency, workload, and
measurement contract differ, these values are not numerically combined with
the current single-host results.

\subsection{Identity Integration and Protocol Interoperability Cases}

Three application-oriented integration cases complement the scalability
experiments: domestic e-commerce, cross-domain data verification, and
AI-for-Science collaboration. They connect registered resources through native
A2A task exchanges and MCP tool calls, using Skill and SDK workflows for
onboarding and discovery, Runtime for controlled identity use, and Sidecar and
Gateway boundaries for protected communication. The cases exercise the
adoption architecture described earlier without replacing the native Agent
protocols or requiring application handlers to manage private keys.

The e-commerce case coordinates six actors across six business interaction
edges for product search, merchant registration-evidence verification,
inventory, logistics, and after-sales decisions. The data-verification case
uses six actors and six edges to compare data digests across domains while
keeping source records within their configured boundary. The AI-for-Science
case uses eight actors and nine edges to connect dataset discovery, experiment
execution, model analysis, and result review. Registration credentials provide
the OAN registration evidence used in these workflows.

The current evidence package passes 21 offline regression checks across the
three case definitions, including build and type checks, protocol-version
compatibility, scenario declarations, and P7--P9 aggregation rules. These
checks confirm that the maintained integration artifacts remain internally
consistent after the protocol and performance changes. They do not constitute
live execution evidence for the three applications. Live validation still
requires the private deployment manifests and complete per-edge evidence
matrices for all three environments.

The cases therefore illustrate how L0 publication and discovery, L1 evidence
checks, and selected L2 identity-bound paths are intended to compose. Their
current status is regression-tested integration design, not a claim about
business throughput, domain validity, or production interoperability.

\section{How to Interpret the Prototype Results}

The evaluation provides evidence for the coupling among lifecycle
governance, Discovery authorization, and trusted invocation under the tested
conditions. The results demonstrate end-to-end execution of the trust workflow and
rejection of tested invalid or unauthorized inputs, while identifying
performance dimensions for further engineering. Together with the public
reference network, these results provide a practical foundation for extending
the architecture across larger and more diverse deployments.

\subsection{Lifecycle Correctness}

Lifecycle correctness means that the same resource identity representation moves
through preparation, credentialing, Root acceptance, package publication,
Discovery synchronization, query, and invocation without losing the identity
binding. This is the most fundamental requirement. A fast system that cannot
preserve this binding is not useful as trust infrastructure.

\subsection{Negative Verification}

Negative verification demonstrates that the system rejects malformed or
unauthorized inputs. These tests are as important as successful-path tests
because trust infrastructure is judged by what it refuses to accept. Invalid
credentials, wrong subjects, unauthorized infrastructure identities, stale
versions, and out-of-scope Discovery cases should fail deterministically.

\subsection{Authorization-Aware Discovery}

Authorization-aware Discovery demonstrates that query matching is not enough.
The returned candidates must also satisfy Root acceptance, current-version
state, and authorized-domain authorization. This result supports OAN's central
claim that Discovery in an Internet of Agents should enforce admissibility rules
alongside semantic relevance.

\subsection{Scalability Characterization}

The progression data localize transient accumulation rather than assigning a
single bottleneck to the whole pipeline. Root and CDN often advance together,
while the largest sampled gaps appear between CDN and the slowest Discovery
target at the 5000-resource scale. Pagination and semantic-index optimization
remain separate concerns because neither per-resource result retrieval nor
semantic convergence is included in the timed completion criterion.

\section{Public Pilot Network and Runtime Transparency}

The reference architecture is also available as a public multi-node network.
This deployment makes the trust path observable through a running setting,
public documentation, and reference interaction surfaces that external readers
can inspect. It demonstrates how the architecture's roles and trust evidence
fit together in an operational environment, complementing the controlled
evaluation with a publicly inspectable reference profile.

These public surfaces do not become protocol authorities. They help readers
observe how Registrar, Root, Discovery, and governance-backed trust evidence fit
together, while the authoritative trust decisions remain in the protocol
objects, signatures, governance state, Root acceptance, and Discovery
verification path.

\begin{figure}[!ht]
\centering
\includegraphics[width=0.47\textwidth]{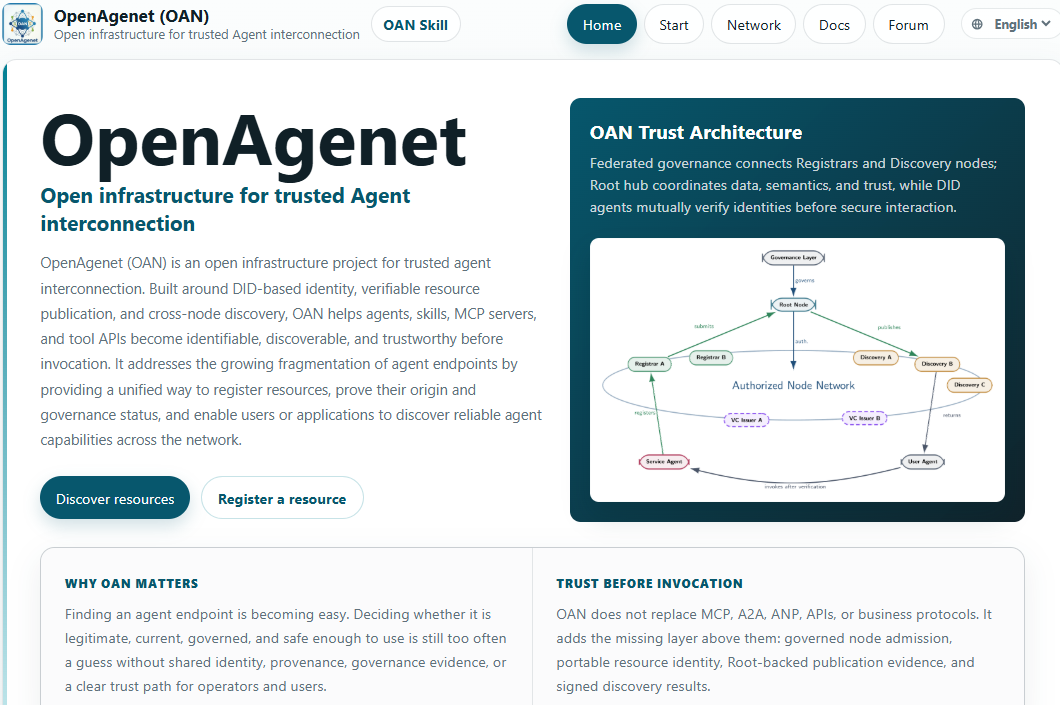}
\caption{Public interaction surface for OAN. The surface illustrates how
registration, discovery, network transparency, and documentation can be exposed
without becoming independent trust authorities.}
\label{fig:wp-public-portal}
\end{figure}

Figure~\ref{fig:wp-public-portal} shows the public interaction surface through
which these roles can be inspected.

\subsection{Multi-Node Pilot Profile}

A multi-node deployment combines Registrar and Discovery nodes under shared
governance rules. Each node follows the same admission, authorization-domain,
distribution, and Discovery semantics. Additional Registrars accept resources
within their authorized domains; additional Discovery nodes synchronize
accepted packages and expose resources within their authorized domains. OAN
can therefore grow by adding operators while preserving the meaning of
registration, Root acceptance, package publication, and Discovery authorization.

A newly authorized Discovery node backfills existing authorized resources from
Root or the distribution layer as well as receiving subsequent updates. This
recovery path lets nodes converge toward the same accepted resource set within
their respective scopes.

\subsection{Network Transparency}

Network transparency makes governed infrastructure observable to ecosystem
participants. Public views can expose topology, resource counts, resource-type
distribution, registration and Discovery activity, node participation, and
governance history. These signals help participants understand network growth
and operation.

\begin{figure*}[!ht]
\centering
\includegraphics[width=0.77\textwidth]{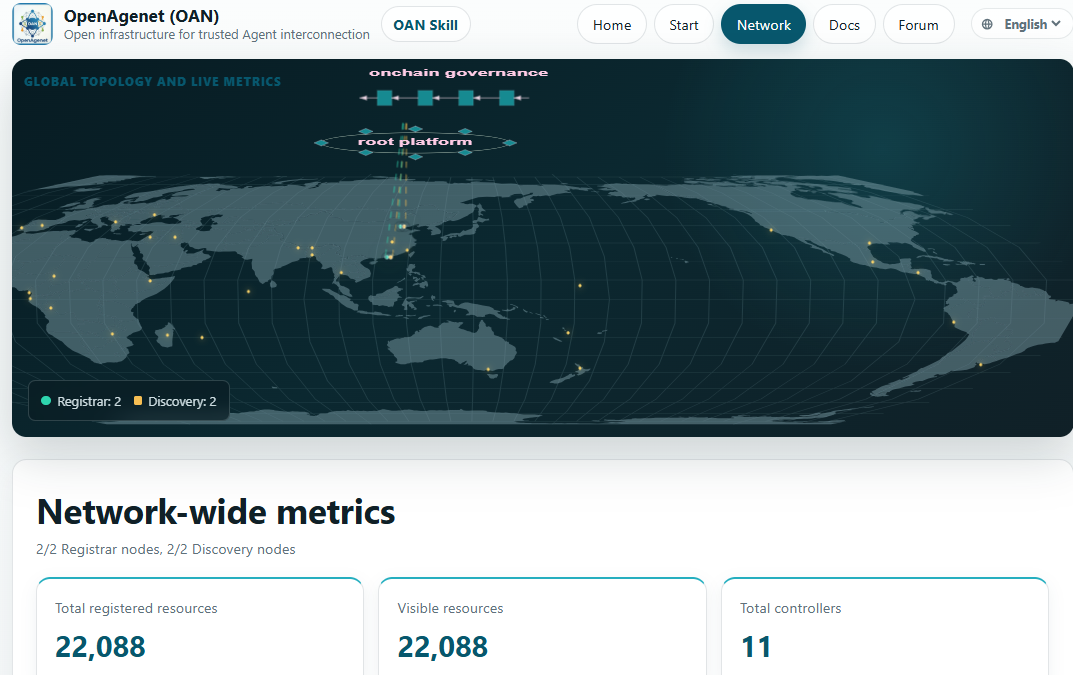}
\caption{Public Network topology view. The view presents a human-readable
snapshot of OAN node participation and topology without exposing internal
deployment details.}
\label{fig:wp-network-map}
\end{figure*}

Network transparency must remain separate from protocol authority. Public
statistics can describe resource counts, resource categories, node-level
activity, and governance-event summaries. They should not expose private logs,
operator secrets, internal service details, or business payloads. They also
should not be treated as governance state or as formal performance evidence.
The authoritative trust path remains the combination of governance state, Root
acceptance, Root-issued evidence, package verification, and Discovery response
verification.

\begin{figure}[!ht]
\centering
\includegraphics[width=0.47\textwidth]{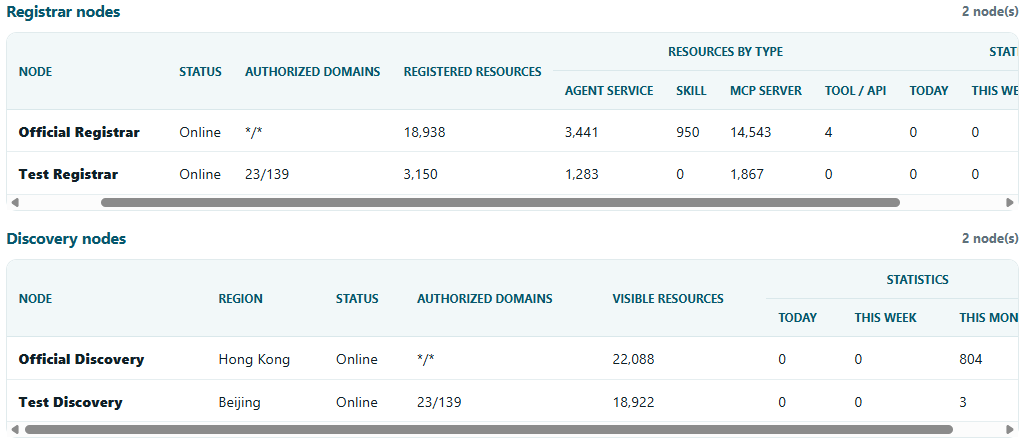}
\caption{Node-level public metrics. Public metrics make Registrar and
Discovery participation visible while keeping protocol trust and operator
secrets outside public presentation surfaces.}
\label{fig:wp-node-metrics}
\end{figure}

Figures~\ref{fig:wp-network-map} and \ref{fig:wp-node-metrics} illustrate the
topology and node-level views that support this human-facing transparency.

\section{Standardization Considerations}

OAN provides concrete interoperability surfaces for standardization. Its
approach prioritizes shared trust semantics, verifiable objects, and conformance
rules while preserving freedom in implementation and deployment. This creates
a practical basis for contributions to domestic and international standards
for Agent identity and trusted interconnection.

\subsection{Candidate Standard Surfaces}

Candidate surfaces include Agent identity representation profiles,
registration credential claims, signed upstream envelopes, Root-verified
package metadata, Discovery response proof format, trusted invocation envelope,
authorized-domain authorization semantics, and conformance test vectors. These
surfaces are useful because they are observable at interoperability boundaries.
They should also include the \texttt{did:oan} identifier profile, DID Document
proof profile, OAN Identity handling conventions for local private material,
controller authorization proof format, external identifier representation, and
the public-key discovery semantics needed by applications that verify
controller or resource login signatures.

OAN's standardization surface should also be understood in relation to wider
Agent identity and governance work. National and industry specifications for
Agent identity can define important identity-code and identity-management
conventions. OAN can complement such work by defining how resource identities
are registered, accepted, packaged, discovered, and verified across nodes.
Similarly, domestic and industry Agent interconnection proposals can be mapped
into OAN when their identity objects, aliases, or governance facts are
represented inside OAN-compatible resource metadata.

The governance layer should remain chain-neutral at the architectural level.
Public blockchains can carry governance evidence for open ecosystems, while
consortium chains can serve regulated or industry-specific deployments.
Off-chain indexers can translate those low-frequency governance facts into
runtime authorization state consumed by Root and infrastructure nodes. This
keeps high-frequency registration and discovery off the governance chain while
preserving traceable authorization history.

\subsection{What Should Remain Flexible}

Operator review process, user interface, database backend, local indexing
strategy, ranking algorithm, business workflow, and domain-specific Agent
protocols should remain flexible. Standardizing these too early would reduce
adoption. OAN should standardize the trust path and let ecosystem participants
innovate above it.

\subsection{Conformance Before Branding}

OAN compatibility depends on consistent verification decisions at
interoperability boundaries. Conformance tests cover object canonicalization,
signature verification, credential proofs, Root packages, Discovery
authorization, current-version handling, and replay rejection. They also cover
DID Document proofs, OAN Identity generation and import, controller
authorization, external identifier handling, and public-key discovery for
application authentication.

\section{Implementation Guidance}

\subsection{Start with the Trust Path}

The shared trust path provides the foundation for independent implementations:
identity creation, DID control proof, registration credential issuance, Root
acceptance and packaging, Discovery synchronization, signed query responses,
and verified invocation. Stable evidence bindings across this path allow
implementations to improve usability and search quality independently.

\subsection{Keep Business Traffic Out of Root}

Root should not become a universal business proxy. It should govern identity,
infrastructure authorization, and package publication. Business traffic between
Agents should be handled by Agent protocols and application systems, with OAN
verification used at connection boundaries.

\subsection{Treat Custom Tags Carefully}

Custom tags are useful for search and domain adaptation. They are also risky if
they accidentally bypass governance. Implementations should treat canonical
authorized domains as the authorization basis and custom tags as refinement
only.

\subsection{Design for Failure}

Registrars may be unavailable, Root publication may lag, CDN synchronization may
fail, Discovery may be stale, and Service Agents may reject invocation
envelopes. A production implementation should surface these states clearly.
Operators need to know whether a missing candidate is caused by registration
failure, Root rejection, publication delay, Discovery authorization, query
mismatch, or runtime verification failure.

\section{Use Cases}

\subsection{Trusted Public-Service Agent Directory}

A public-service operator may need to expose Agents for policy consultation,
service navigation, document preparation, or process guidance. OAN allows such
Agents to be registered, governed, and discovered under a shared trust domain.
Users and platforms can verify that a listed Agent is Root-accepted and current.

\subsection{Industrial Agent Cooperation}

Industrial Agents may operate across manufacturing, logistics, quality control,
energy management, and supply-chain workflows. Authorized-domain Discovery can
separate governed exposure scopes while allowing approved cross-domain
cooperation. Existing industrial protocols can remain in place while OAN
provides identity and discovery trust.

\subsection{Research and Testbed Collaboration}

Universities, laboratories, and standards groups can use OAN as a testbed for
Agent identity, discovery, governance, and interoperability. The role
separation makes it possible to test multiple Registrar and Discovery operators
without rewriting the whole stack.

\subsection{Enterprise Agent Gateway}

An enterprise can register internal or partner-facing Agents through a
Registrar, expose selected capabilities through a Discovery node, and require
OAN verification before external Agents access internal workflows. This pattern
can coexist with enterprise identity and access management (IAM), API
gateways, and data-governance systems.

\section{Cooperation with Other Organizations}

OAN is useful for cooperation because it does not require partners to abandon
their current systems. A partner can keep its own Agent framework, MCP tools,
A2A task logic, enterprise workflow engine, or domain API while integrating OAN
for identity and discovery.

Practical cooperation can begin with a small set of tasks:

\begin{itemize}[leftmargin=*]
  \item map the partner's Agent, Skill, MCP Server, or Tool/API resources to
  OAN resource types and authorized domains;
  \item create OAN-compatible \texttt{did:oan} documents for selected
  resources;
  \item register selected resources through a trial Registrar;
  \item expose them through an authorized Discovery node;
  \item add OAN verification before MCP, A2A, REST, or domain-protocol
  sessions;
  \item define conformance checks and shared test cases.
\end{itemize}

This lets partners collaborate on the trust layer first, without forcing early
agreement on every application protocol detail.

For organizations that already use ADS, ANS, MCP registries, Skill
marketplaces, the preferred cooperation model is not replacement. It is
wrapping and mapping. Existing records can be mapped into OAN ResourcePackages;
existing names can be represented as aliases; existing schema objects can be
embedded or referenced as metadata; and existing runtimes can keep executing
tasks after OAN has completed pre-connection verification.

The open-source path is also part of cooperation. A partner should be able to
start from documentation, inspect the protocol and reference services, use SDKs
for Agent-side verification, register a small number of resources, and then
decide whether it needs to operate infrastructure nodes. This keeps the first
step small while preserving a route toward deeper participation. It also makes
standardization discussion more concrete: participants can debate object
models, trust evidence, conformance checks, and governance vocabulary against a
running reference stack instead of only comparing abstract diagrams.

\section{Stakeholder Value}

OAN's role separation supports different forms of participation. Resource
developers publish verifiable identities, infrastructure operators provide
governed services, and relying organizations apply the resulting evidence to
their own interaction policies.

\subsection{Resource Developers}

Resource developers can publish Agent Services, Skills, MCP Servers, or
Tool/API endpoints as accepted identities and discoverable packages. The
published material binds identity, endpoint or artifact references,
capabilities, credentials, and verification methods in a form that relying
systems can inspect, reducing dependence on private documents and manually
exchanged endpoints.

For developers, OAN also reduces repeated integration work. Once a resource can
produce OAN-compatible identity material and signed responses, it can be made
visible through different Discovery nodes and can be invoked by different User
Agents. This is especially valuable for small teams that cannot negotiate a
custom trust integration with every potential partner.

\subsection{Infrastructure Operators}

OAN separates Root, Registrar, Discovery, and CDN responsibilities so that an
operator can participate without owning the entire ecosystem. A Registrar
operator can focus on onboarding and
evidence collection. A Discovery operator can focus on search, indexing, and
domain-specific query quality. A Root operator can focus on governance-state
consumption, policy enforcement, audit, credential issuance, and
accepted-version authority. A CDN operator can focus on package availability.

This role separation also allows multiple organizations to participate in a
pilot network without collapsing into one centralized service provider.

\subsection{Enterprise Users}

For enterprise use, OAN supplies evidence about an external Agent's identity,
the governance of its discoverability, and the caller and target of an
invocation. This pre-connection evidence does not require the enterprise to
expose internal business systems through OAN.
The enterprise can keep its existing access-control, auditing, and data
governance mechanisms while using OAN as a pre-connection trust layer.

\subsection{Public-Service and Industry Platforms}

Public-service and industry platforms often require transparent governance and
traceability. OAN's Root acceptance records, Registrar credential path, and
Discovery authorization model provide a concrete basis for explaining why an
Agent appears in a directory. This is important when Agents represent public
service functions, regulated-domain guidance, or industrial workflows where
uncontrolled exposure can create real operational risk.

\subsection{Standards and Research Communities}

OAN provides a concrete implementation target for discussions about Agent
identity, Agent discovery, and Agent-to-Agent trust. Its implementation,
tests, examples, and
experimental results can support standards discussion, interoperability tests,
and academic evaluation.

\section{Governance Operations}

Governance operations translate the trust model into infrastructure lifecycle
decisions, taxonomy maintenance, and incident response. The following
operations distinguish governance decisions from Root-side enforcement.

\subsection{Registrar Authorization Lifecycle}

A Registrar authorization lifecycle includes application, review, activation,
monitoring, suspension, and revocation. During application, an operator
describes its intended authorized domains, security posture, key management
plan, and operational contact. During activation, the governance operator
records the Registrar DID, lifecycle state, and approved authorized domains in
the bulletin. During monitoring, Root
and ecosystem maintainers may
inspect submission quality, rejection patterns, abuse reports, or operational
health. Suspension and revocation should be reflected in Root state so that
future submissions are rejected.

\subsection{Discovery Authorization Lifecycle}

Discovery authorization is domain-sensitive. A Discovery node should receive a
defined authorized-domain scope. That scope can be expanded, reduced, paused,
or revoked. Because Discovery controls exposure, its authorization should be
treated as a public trust fact. The on-chain governance layer allows Discovery
authorization changes to be sensed by infrastructure nodes through events. A
relying Agent should be able to know whether a Discovery node is currently
authorized and which domains it may expose through verified Discovery evidence.

\subsection{Semantic Tag Tree Governance}

The authorization-domain taxonomy and semantic tag tree are not only technical
indexes. They are governance artifacts. Adding, renaming, splitting, or merging
domains can affect which resources a Registrar may admit and which resources
become visible through which Discovery nodes. Adding or reorganizing semantic
tags can affect how resources are described during registration, how tags are
recommended to publishers, and how Discovery maps user intent to resource
candidates. OAN therefore treats domain-taxonomy and semantic-tag changes as
governance operations. A production ecosystem should keep versions, change
records, migration guidance, and compatibility rules for authorization domains
and semantic tags.

\subsection{Incident Response}

An Agent may be compromised, a Registrar may issue invalid credentials, a
Discovery node may expose out-of-scope identities, or a package distribution
path may be attacked. OAN should support incident response through revocation,
state updates, publication of new Root facts, package re-synchronization,
Discovery re-indexing, and relying-party verification. The goal is not to
prevent every incident, but to make incidents observable and recoverable.

\section{Data and Evidence Model}

OAN separates several kinds of data that are often mixed together in simpler
directory systems.

\subsection{Public Identity Data}

Public identity data includes resource DID, verification methods, service
endpoints or artifact references, capability tags, public metadata,
controller declarations, external identifiers, and Root-verified package
references. This data is meant to be used for discovery and verification. It
should be sufficient for a relying party to decide whether a candidate is a
valid OAN identity and how to approach it. Private key material remains outside
the public identity data model. An OAN Identity may hold private signing
material locally, but a DID Document, VC, Root package, Discovery response, or
public identity directory entry should expose only public verification
material and verifiable metadata.

\subsection{Controller and External-Identifier Evidence}

Controller evidence explains who may authorize a registration operation or
control a resource relationship. A self-controlled resource binds its own DID
and keys directly. A dedicated controller or external controller uses a
separate controller identity and explicit authorization evidence. This
separation allows applications to verify a controller signature without
treating a controller string as sufficient proof of authority.

External-identifier evidence has a different purpose. It maps an OAN resource
to a native platform or industry identifier. Such a mapping improves
interoperability and traceability, but it is not by itself proof that the
external platform has endorsed the OAN resource. Relying parties should
interpret it as signed metadata bound to the resource DID Document. External
ownership or endorsement requires separate evidence from the external platform
or an authorized credential issuer.

\subsection{Governance Evidence}

Governance evidence may include operator applications, review notes, enterprise
approval documents, or offline contact verification. This evidence may be
important for governance, but it does not always belong in public packages. OAN
therefore allows the protocol-facing trust path to rely on signed credentials
and Root records while leaving private evidence in controlled operational
systems.

\subsection{Runtime Invocation Evidence}

Runtime invocation evidence includes signed request envelopes, nonces,
timestamps, selected-service provenance, and signed responses. This evidence is
used to verify a specific interaction attempt. It is different from identity
admission evidence. Keeping the two separate helps avoid requiring Root to
observe every business request.

\subsection{Audit Evidence}

Audit evidence links the lifecycle together: which Registrar submitted a
representation, which credential was used, which Root record accepted it, which
package was published, which Discovery node indexed it, and which response was
returned. OAN's architecture is designed so that this chain can be reconstructed
without relying on informal claims.

\section{Interoperability Strategy}

OAN's interoperability strategy has four layers.

\subsection{Object Compatibility}

The first layer is object compatibility. DID-style documents, credential-like
claims, signed envelopes, Root packages, and Discovery responses should have
stable schemas and canonical verification rules. SDKs should help developers
produce and consume these objects without hand-implementing every cryptographic
detail. OAN Identity handling belongs at this layer as a local-key convention:
implementations should never confuse local private signing material with public
DID Documents, VCs, or Discovery responses.

\subsection{Protocol Adapter Compatibility}

The second layer is adapter compatibility. MCP, A2A, ANP-like protocols, REST
APIs, and domain protocols can use OAN verification before opening a session or
executing a request. The adapter should translate OAN trust context into the
local protocol's authentication, authorization, or metadata model. For example,
an application login adapter can map an OAN Identity challenge signature and a
Root-accepted public-key projection into the application's session model
without making the application itself a DID resolver or VC issuer.

\subsection{Operational Compatibility}

The third layer is operational compatibility. Operators need deployment,
configuration, logging, and monitoring conventions that make independent
implementations comparable. Without operational compatibility, a protocol can
look interoperable on paper while failing in real deployments.

\subsection{Governance Compatibility}

The fourth layer is governance compatibility. Different organizations may use
different review processes, but they need shared meanings for accepted,
revoked, authorized, current, and discoverable states. OAN's governance-layer
lifecycle events, Root-issued infrastructure authorization VCs, Root records,
and authorization-domain model provide a basis for this shared vocabulary.

\section{Adoption Path}

OAN adoption can be staged. A staged path lowers integration risk and allows
partners to choose a depth of participation.
The stages are not mandatory in a fixed order. A partner may begin as an SDK
consumer, operate a Discovery node for one capability domain, run a Registrar
for a governed community, or build an adapter that brings OAN verification into
an existing MCP, A2A, ANP-like, or domain-specific protocol. This flexibility
is important for cooperation with organizations that already have operational
systems and cannot replace them in one step.

\begin{table}[t]
\centering
\caption{Representative Adoption Depths}
\label{tab:adoption-depths}
\begin{tabular}{@{}lll@{}}
\toprule
Depth & Suitable participants & Typical action \\
\midrule
Observe & reviewers, visitors & inspect portal, Network, Docs \\
Publish & resource providers & register resources \\
Discover & users, applications & query trusted candidates \\
Integrate & developers & use SDKs and DID method docs \\
Operate & node operators & prepare Registrar or Discovery nodes \\
\bottomrule
\end{tabular}
\end{table}

Table~\ref{tab:adoption-depths} shows that OAN can be adopted at different
depths. A visitor can inspect the public network. A resource provider can
register an Agent Service, Skill, MCP Server, or Tool/API endpoint. A developer
can use SDKs and the \texttt{did:oan} method specification. An infrastructure
operator can prepare a Registrar or Discovery node under governance admission.
This layered path is important because open infrastructure must support both
lightweight evaluation and deeper operational participation. These participation
stages are orthogonal to the L0--L3 integration depths: a partner may publish
at L0, verify at L1, or enforce L2 through a Runtime, Sidecar, or Gateway
without operating a Registrar or Discovery node. Conversely, operating an
infrastructure node does not automatically make its own Agent traffic an L2
protected interaction.

Public interaction surfaces and SDKs serve different adoption moments. The
public surfaces make the network understandable and testable. SDKs let
applications generate OAN Identity material, verify DID Documents, verify
credentials, consume Discovery responses, and create signed invocation evidence
without reimplementing the trust path. These adoption mechanisms should remain
consistent with the same underlying trust model rather than becoming separate
sources of authority.

\subsection{Stage 1: Identity Publication}

In the first stage, an organization creates OAN-compatible identity documents
for a small number of Agents. The goal is to make identity material explicit:
DID, verification methods, endpoints, capabilities, and basic metadata.

\subsection{Stage 2: Governed Registration}

In the second stage, Agents are registered through Registrar and accepted by
Root. The organization learns the evidence path, subject-control proof, and
Root package publication workflow.

\subsection{Stage 3: Trusted Discovery}

In the third stage, an authorized Discovery node indexes accepted packages and
returns signed discovery responses. Partner systems can start using OAN
Discovery instead of manually curated endpoint lists.

\subsection{Stage 4: Trusted Invocation}

In the fourth stage, User Agents and Service Agents use signed invocation
envelopes. The integration now covers the full pre-connection path from
identity to first verified request.

\subsection{Stage 5: Protocol Integration}

In the fifth stage, OAN verification is embedded into MCP, A2A, ANP-like, or
domain-specific protocol flows. At this point, OAN becomes part of routine
Agent interconnection rather than a separate demonstration.

\section{Economic and Ecosystem Considerations}

Open infrastructure must avoid unnecessary lock-in. OAN's role separation
supports different ecosystem models. A public trust domain may operate one Root
with multiple independent Registrars and Discovery nodes. An enterprise may
operate a private Root for internal and partner Agents. A research testbed may
allow experimental roots and compare policy designs. A standards community may
use OAN object models and conformance tests without mandating a single public
operator.

The ecosystem should also allow specialization. Some Discovery nodes may
optimize for low-latency industrial search. Others may optimize for public
service transparency. Some Registrars may specialize in enterprise Agents,
while others focus on research or developer communities. OAN gives these roles
shared trust semantics while allowing different service models above them.

\section{Limitations}

The present architecture and evidence leave four material limitations.

\subsection{Single Trust Domain Scope}

The current architecture operates within one Root trust domain. It does not yet
specify or validate how independently governed domains recognize one another's
Root records, governance policies, or Discovery exposure rules. Multi-root
federation therefore remains a protocol and evaluation task.

\subsection{Semantic Quality}

OAN verifies identity and governance state, not the truth of every capability
claim or the quality and safety of resource behavior. Those properties require
separate testing, reputation, certification, runtime monitoring, or domain
policy. The reported benchmarks also exclude semantic-index convergence.

\subsection{Privacy}

The current Discovery model primarily handles publishable resource evidence.
Private queries, selective disclosure, and confidential capability matching
are not yet defined or evaluated for deployments with stronger
data-minimization requirements.

\subsection{Production Hardening}

The reference implementation establishes the main architecture and integration
paths, but production deployments still require deployment-specific key
custody, rate limits, monitoring, backup, disaster recovery, and operational
security controls.

\section{Reference Implementation Profile}

The OAN reference implementation follows the role separation described in the
architecture. Root provides acceptance, proof issuance, package publication,
distribution coordination, and trust-domain services. Registrar provides
governed onboarding, DID Document validation, registration review, and Root
submission. Discovery provides package synchronization, structured indexing,
derived semantic indexing, scoped query, and signed Discovery results. The
Trust Indexer converts low-frequency governance events into
runtime-readable infrastructure authorization state. Client libraries help
applications create OAN Identity material, produce signed envelopes, verify
DID Documents and VCs, and consume Discovery responses.

The reference profile separates public protocol roles from implementation
choices such as repository layout and deployment procedure. Partners can
adopt the parts they need: resource publishers may
start with identity generation and registration, application developers may use
client-side verification, and infrastructure operators may run Registrar or
Discovery nodes under governance admission. This lets OAN grow as open
infrastructure rather than a monolithic application.

\section{Roadmap}

\subsection{Near-Term Consolidation and Extension}

The current baseline already includes governed registration and Discovery,
multiple Registrar and Discovery roles, public network and governance views,
OAN Identity and controller-aware flows, external-identifier handling, and
reference integration paths for MCP and A2A. The near-term roadmap builds on
this foundation by consolidating deployment profiles and extending
interoperability coverage. The Root reference implementation now uses a durable
SQLite nonce store and atomically maintains current controller state; remaining
priorities include key custody, rate limiting, observability, recovery, and
operational governance; broader MCP and A2A conformance for asynchronous tasks, streaming,
cancellation, callbacks, and complete resource or artifact lifecycles; and
richer semantic Discovery, backfill, result inspectability, independent
conformance vectors, and operator admission documentation. Ecosystem
construction will also connect OAN with
more third-party credential-issuing infrastructure, so that reputation ratings,
capability certifications, industry attestations, and external trust bridges
can be used alongside OAN registration evidence. These credentials should enrich
resource evaluation and cross-ecosystem recognition without replacing OAN's
resource identity, Root acceptance, or Discovery authorization.

\subsection{Long-Term}

Long-term extensions include multi-root federation, privacy-preserving Discovery,
cross-domain capability governance, independent implementation conformance,
formal security analysis, and broader standardization engagement. These
directions position OAN to grow from a shared trust foundation into widely
interoperable infrastructure for open Agent ecosystems.

\section{Conclusion}

OAN places resource identity, controller authorization, admission evidence,
lifecycle state, and scoped discovery below application protocols. Governance,
Root policy enforcement, Registrar onboarding, CDN distribution, Discovery
exposure, and business invocation remain separate responsibilities. This
separation lets an existing Agent system adopt governed publication and
evidence checking without replacing MCP, A2A, its runtime, or its business
workflow.

The reference implementation demonstrates the common lifecycle for four
resource types and reaches complete Root, CDN, and Discovery counts in all 60
structured benchmark runs. Controller-continuity and negative security gates
exercise the principal authorization boundaries. The 60-run performance matrix
uses one host. A separate three-host experiment confirms physical deployment,
but is not a controlled scale-out comparison; semantic convergence was not
measured, and the three application cases have offline regression evidence
rather than live deployment evidence.

The next steps are independent conformance evidence, live protocol-integration
reports, semantic evaluation with the required runtime, and federation beyond
the present single-Root trust domain.

\section*{Acknowledgment}

The author thanks Jin Jian, Xie Jiagui, Li Bingqi, and Zhu Runkai for their
support.

\nocite{xu2026grail,xu2026darwinnet,qian2024chatdev,huang2025ans,cui2025agentdns,raskar2025nanda}
\bibliographystyle{IEEEtran}
\bibliography{Ref}

\end{document}